\documentclass[aps,showpacs,twocolumn,amsmath,amssymb,nofootinbib]{revtex4-1}
\usepackage{graphicx,epsfig}
\pdfoutput=1
\usepackage{amsfonts}
\usepackage{mathrsfs}
\usepackage{multirow}

\def\Journal#1#2#3#4{{#1} {\bf #2}, #3 (#4)}

\def\ANNP{Ann. Phys. (N.Y.)}
\def\EPJC{{Eur. Phys. J.} C}
\def\NCA{Nuovo Cimento}

\def\PLB{{Phys. Lett.}  B}
\def\PLA{{Phys. Lett.}  A}
\def\PRL{Phys. Rev. Lett.}
\def\PRD{{Phys. Rev.} D}
\def\PR{{Phys. Rev.}}

\def\NAT{{Nature}}

\def\CQG{{Class. Quant. Grav.}}
\def\CMP{{Commun. Math. Phys.}}
\def\JMP{{J. Math. Phys.}}
\def\JPA{{J. Phys.}  A}
\def\JHEP{{J. High E. Phys.}}

\def\IJMPD{{Int. J. Mod. Phys.} D}

\def\ibid{{\it ibid.}}

\newcommand{\be}{\begin{equation}}
\newcommand{\ee}{\end{equation}}
\newcommand{\bea}{\begin{eqnarray}}
\newcommand{\eea}{\end{eqnarray}}

\begin{document}

\title{Obstructions towards a generalization of no-hair theorems: I. Scalar clouds around Kerr black holes}

\date{\today}


\author{Gustavo Garc\'ia}
\email{gustavo.garcia@correo.nucleares.unam.mx} 
 \author{Marcelo Salgado}
\email{marcelo@nucleares.unam.mx} \affiliation{Instituto de Ciencias
Nucleares, Universidad Nacional Aut\'onoma de M\'exico,\\
 A.P. 70-543, M\'exico 04510 D.F., M\'exico}

\begin{abstract}
  We show that the integral method used to prove the no-hair theorem for Black Holes (BH's)  in
  spherically symmetric and static spacetimes within the framework of general relativity 
with matter composed by a complex-valued scalar-field does not lead to 
a straightforward conclusion about the absence of hair in the stationary and rotating (axisymmetric) scenario.
We argue that such a failure can be used to justify in a simple and heuristic way the existence of
non-trivial boson clouds or {\it hair} found numerically by Herdeiro and
Radu~\cite{Herdeiro2014,Herdeiro2015} and analytically by Hod in the test field limit~\cite{Hod2012,Hod2013,Hod2016}. This is due to the presence of a contribution that is negative when
rotation exists which allows for an integral to vanish even when a non-trivial boson hair is present. 
The presence of such a negative contribution that depends on the rotation properties of the BH
is perfectly correlated with the eigenvalue problem associated with the boson-field equation.
Conversely, when the rotation is absent the integral turns to be composed only by non negative (i.e. positive semidefinite) 
terms and thus the only way it can vanish is when the hair is completely absent. 
This analysis poses serious challenges and obstructions towards the elaboration of no-hair theorems 
for more general spacetimes endowed with a BH region even when including matter fields that obey the energy conditions.
Thus {\it rotating} boson stars, if collapsed, may lead indeed to a new type of rotating BH, like the
ones found in~\cite{Herdeiro2014,Herdeiro2015}. In order to achieve this analysis we solve numerically the eigenvalue problem 
for the boson field in the Kerr-BH background by imposing rigorous regularity conditions at the BH horizon for the 
non-extremal case ($0<a<M$) which include the near extremal one in the limit $M\rightarrow a$, 
as well as the small BH limit $M\rightarrow a\rightarrow 0$.
 \end{abstract}

 \maketitle

Wheeler's famous no-hair conjecture establishes that asymptotically
flat black holes (AFBH) are characterized only by three parameters: mass, charge and angular momentum~\cite{ruffini}.
This conjecture is supported by the uniqueness theorems which assert that all regular AFBH solutions 
within the Einstein-Maxwell system that are stationary and axially symmetric are contained within the 
Kerr-Newman family which is fully described by those three parameters~\cite{Uniqueness} (see~\cite{Heusler1996} for a review). There are, however, some counterexamples to this conjecture 
when other field theories are taken into account, notably, when including non-abelian 
gauge fields like in the  Einstein-Yang-Mills system~\cite{eym} or when the energy-momentum tensor (EMT) of some field theories 
do not obey the positive-energy conditions, like in 
the Einstein-Higgs system with a non-positive semidefinite scalar-field potential $V(\phi)$~\cite{Nucamendi2003,Anabalon2012}. 
Indeed the so-called no-hair theorems for matter fields composed by a real or complex-valued (boson) scalar 
fields in static and spherically symmetric spacetimes within the framework of general relativity apply
provided the potentials obey the energy conditions, 
namely, the weak energy condition (WEC) $V_\phi(\phi)\geq 0$, $V_\Psi(\Psi^*\Psi)\geq 0$
~\cite{Sudarsky1995,Heusler1992,Bekenstein1995} (see \cite{Herdeiro2015c} for a review). One way to prove these theorems is to
use the conservation equation for the EMT of the matter fields. This equation 
together with the AF condition and the WEC leads to the conclusion that the only possible field configurations
are the trivial ones $\phi(r)=0$ and $\Psi(t,r)\equiv 0$~\cite{Sudarsky1995,Bekenstein1995}. 
Other proofs use the Klein-Gordon equation which, by some manipulations, 
can be integrated in suitable but arbitrary region of the domain of outer communications of the BH (see Section~\ref{sec:clouds})
leading to a vanishing integral whose integrand is non-negative if the potential is convex. This integral can thus be satisfied if and only if 
the scalar field vanishes identically~\cite{Bekenstein1972}.

In this paper we show that if one includes rotation, the integral used to prove the no-hair theorem for a complex-valued 
boson field within a spherically symmetric and static spacetime does not lead to the same conclusion for a stationary and axisymmetric 
AFBH (namely, the boson field does {\it not} necessarily vanish) since the rotation
produces terms in the integral that are not positive semidefinite. Thus, {\it a priori}, the inclusion of less symmetries in the problem 
generate obstructions to extend the no-hair theorems. Surprisingly, it turns out, that these obstructions lead to 
actual counterexamples to the no-hair conjecture in the rotating case even when the WEC is satisfied. 
The first in showing some evidence about the existence of non-trivial complex-valued field configurations was Hod~\cite{Hod2012,Hod2013}, 
who assumed an {\it extremal} and {\it near} extremal  Kerr BH backgrounds. By solving the corresponding eigenvalue problem Hod proved that 
non-trivial configurations exist for the boson field. These configurations were termed {\it clouds}. 
Later, Herdeiro and Radu~\cite{Herdeiro2014,Herdeiro2015} reached the same conclusions numerically for general 
Kerr BH configurations that were not necessarily extremal, and 
Hod~\cite{Hod2016} provided analytic expressions for the spectrum in the large limit $M\mu$ (cf. Sections~\ref{sec:clouds} 
and~\ref{sec:numerics}). Benone {\it et al.}~\cite{Benone2014} generalized these results for a Kerr-Newman background. 
Herdeiro and Radu~\cite{Herdeiro2014,Herdeiro2015} 
also analyzed the scenario where the background is not fixed and solved numerically the full fledged Einstein-Boson field system 
selfconsistently finding the existence of stationary and rotating regular AFBH's endowed with 
boson hair. They also showed that in the limit where the BH's horizon shrink to zero, the resulting configuration corresponds to a 
globally regular rotating boson star. Further analysis by the same authors together with other collaborators 
extended those results by including a self-interacting boson field~\cite{Herdeiro2015b} and a charged boson field, i.e., for the 
latter study they solved numerically the Einstein-Maxwell-(un)charged-boson field system~\cite{Herdeiro2016}.
Previously Hod~\cite{Hod2014} had performed a study of charged clouds in the Kerr-Newman background for the near extremal case.

In the present paper we focus only on the analysis of boson clouds within a fixed Kerr background. In a forthcoming work we plan to analyze a 
similar scenario within the (charged) Kerr-Newman BH. In order to accomplish our goal we solve the 
corresponding eigenvalue problem and show that when non-trivial boson-field configurations, i.e. boson clouds,
exist negative contributions
within an integral identity emerge allowing for this to vanish without the requirement for a trivial field configuration to be present. The existence of these clouds depends on solving an eigenvalue problem that leads to a
{\it quantized} angular momentum $a=J/M$ for the Kerr BH. In the limit of zero angular momentum the BH's horizon shrink to
zero and the cloud's frequency $\omega$ saturates the mass $\mu$ of the boson field. In that limit the boson field do not 
fall-off asymptotically in the form of a Yukawa fashion but just $\sim 1/r$ and the angular velocity of the BH becomes maximal.
We argue that in the exact scenario where $r_H=0=a=M$, the BH ``disappears'' and the spacetime becomes Minkowski 
and the clouds become singular at the origin. The conclusion is that regular boson clouds exist only due to the
rotation of the Kerr BH and the existence of such clouds presumably require that the BH angular velocity $\Omega_H$
be bounded between the extremal case and the maximum frequency: $\Omega_H^{\rm ext}<\Omega_H< \mu/m$, where
$m$ is an integer (a {\it quantum number}) associated with the angular dependency of the boson field. Thus, {\it regular} clouds cannot exist 
in the absence of rotation, in agreement with the no-hair theorems for spherically symmetric spacetimes, and in the absence of 
a BH (i.e. in Minkowski spacetime).

One last comment of paramount importance is in order. In the past, BH solutions with non trivial hair 
with singular behavior at the BH horizon have been reported~\cite{BMBB}. Since such solutions are not genuinely regular their 
significance was the subject of debate~\cite{Sudarsky1998}. Thus, in order to avoid such kind of drawback it is {\it essential} 
to impose suitable regularity conditions on the matter fields at the BH horizon. For our numerical analysis we 
impose those kind of conditions to the non-extremal solutions which include the near extremal ones in the limit
$M\rightarrow a$. The latter solutions present radial functions with unbounded gradients at $r_H=M=a$ (cf. Sec.~\ref{sec:numerics}),
notwithstanding, we argue that, given the type of the 
divergence, the invariant scalars formed from derivatives of the fields may remain finite at $r_H=a$.
In a future work we will analyze if extremal cloud solutions with bounded gradients of the field at $r_H=a$ exist or not.

\section{The cloud scenario}
\label{sec:clouds}

We assume a Kerr BH background with a metric given in terms of the Boyer-Lindquist coordinates
\footnote{Notation may vary from textbooks and monographs. For instance, Wald~\cite{Wald1984} uses 
$\Sigma$ instead of $\rho^2$ which is used by Chandrasekhar~\cite{Chandrasehkar1992}.}:
\begin{eqnarray}
\label{Kerr} 
\nonumber && ds^2 = -\left(\dfrac{\Delta - a^2\sin^{2}\theta}{\rho^2}\right)dt^2 - \dfrac{2a\sin^2\theta\left(r^2 + a^2 - \Delta\right)}{\rho^2}dtd\varphi \\  
& & + \dfrac{\rho^2}{\Delta}dr^2 + \rho^2d\theta^2 + \left(\dfrac{\left(r^2 + a^2\right)^2 - \Delta a^2\sin^{2}\theta}{\rho^2}\right)\sin^2\theta d\varphi^2, 
\end{eqnarray}
where
\begin{eqnarray}
\Delta &=& r^2 - 2Mr + a^2 \;,\\
\rho^2   &=& r^2 + a^2\cos^2\theta \;,
\end{eqnarray}
$M$ and $a$ are the {\it mass} and {\it angular momentum} (per mass unit) associated with the Kerr BH. 

We consider a complex-valued massive and free scalar field $\Psi$ which has the following energy-momentum tensor\footnote{If one 
redefines $\Psi=\sqrt{2}{\tilde \Psi}$ one recover the usual parametrization of the EMT used in \cite{Herdeiro2014,Herdeiro2015}.}
\begin{eqnarray}
\label{Tab}
T_{ab} &=& \frac{1}{2}\Big[\nabla_a\Psi^* \nabla_b \Psi+ \nabla_b\Psi^* \nabla_a\Psi\Big] \nonumber \\
&-& g_{ab}\Big[\frac{1}{2}g^{cd}\nabla_c\Psi^* \nabla_d\Psi^* + V(\Psi^*\Psi)\Big]  \;, \\
V(\Phi^*\Phi) &=& \frac{1}{2}\mu^2 \Psi^*\Psi \;,
\end{eqnarray}
where $\mu$ is the mass associated with $\Psi$. The {\it cloud} solution that we aim to find corresponds to
a boson field that does not backreact on the Kerr background, and thus, the field is considered to be a {\it test field} 
in the sense that the above EMT does not contribute as source to the Einstein field equations. Nonetheless, the 
field equation for $\Psi$ that we write below is analyzed in the curved spacetime associated with the fixed Kerr background.

The complex boson field $\Psi$ obeys the Klein-Gordon equation
\begin{equation}
\label{KG}
\Box \Psi= \mu^2 \Psi \;,
\end{equation}
where $\Box= g^{ab}\nabla_a \nabla_b$ is the covariant d'Alambertian operator, and $g_{ab}$ corresponds to 
the Kerr metric (\ref{Kerr}). In order to find ``bound state'' 
solutions we assume $\Psi(t,r,\theta,\varphi)$ to have a time and angle dependency in the following form
\begin{equation}
\label{Psians}
\Psi(t,r,\theta,\varphi)= e^{\imath (-\omega t + m \varphi)} \phi(r,\theta) \;,
\end{equation}
where $\phi(r,\theta)$ is real valued and $m$ is a non-zero integer. The harmonic dependence 
of the boson field is such that the EMT respects the symmetries of the underlying spacetime. The idea is that this kind of test field 
is just an approximation of a more realistic solution where the field is allowed to backreact on the 
spacetime (cf. \cite{Herdeiro2014,Herdeiro2015}).

The angular velocity of the Kerr BH is given in terms of the BH {\it horizon} located at $r_H$ and the angular momentum $a$ by
\footnote{\label{foot:BH3+1}
In the 3+1 formalism of general relativity (cf. Section~\ref{sec:eingenprob}), the BH's angular velocity  $\Omega_H$ can be defined as the angular velocity of the {\it normal observer} to the hypersurfaces $t=const$ at the horizon~\cite{Wald1984}. 
This angular velocity is no other but the {\it shift-vector} component 
$N^{\varphi}_H$ \cite{Wald1984}. Alternatively,
the timelike and axial Killing fields $\xi^a= (\partial/\partial t)^a$ and $\eta^a= (\partial/\partial \varphi)^a$ are orthogonal 
to $\chi^a := \xi^a + \Omega_H \eta^a$ at the horizon [cf. Eq.~(\ref{chi})] leading to $\Omega_H= - \xi_a \xi^a/(\xi_a\eta^a)|_H= - \eta_a \xi^a/(\eta_a\eta^a)|_H$
\cite{Heusler1996} and to the expression (\ref{OmH}). In turn, the (Killing) horizon is defined to be as the region where $\chi^a \chi_a|_H=0$ which 
leads to Eq.(\ref{rH}). In fact $\chi^a \chi_a|_H= -N^2|_H$ where $N$ is the {\it lapse} function~\cite{Wald1984}, therefore the horizon corresponds to the 
spacetime region where the lapse vanishes. At the horizon the normal observer has 4-velocity $n^a|_H=\chi^a_H/N_H$ and satisfies 
$n^an_a=-1$. For the Kerr metric the normal observer coincides with the so-called Zero Angular Momentum Observer 
(ZAMO) since $n^a= \frac{1}{N}\left(\xi^a + N^\varphi \eta^a \right)$ and $L_{ZAMO}= n^a \eta_a\equiv 0$.}

\begin{equation}
\label{OmH}
\Omega_H=  \frac{a}{r_H^2 + a^2}= \frac{a}{2Mr_H} \;,
\end{equation}
where $r_H=r_+$ is the largest root of the algebraic equation
\begin{equation}
\label{rH}
\Delta_{\pm} =r^2_{\pm} - 2Mr_{\pm} + a^2=0 \;,
\end{equation}
that is
\begin{equation}
r_H=r_+= M+\sqrt{M^2-a^2}\;.
\end{equation}
The other root is given by
\begin{equation}
r_{-}= M-\sqrt{M^2-a^2}\;.
\end{equation}
The existence of a Kerr BH requires $|a|\leq M$.

So we can write
\begin{equation}
\label{Delta}
\Delta= (r-r_H)(r-r_{-})\;.
\end{equation}
Moreover, there is also the following relationship between the two horizons:
\begin{equation}
  \label{r-}
r_{-}= \frac{a^2}{r_H} \;.
\end{equation}

So given $r_H$ and $a$ one can compute $r_{-}$ from Eq.~(\ref{r-}) and $M$ from
\begin{equation}
\label{BHmass}
M= \frac{r_H^2 + a^2}{2r_H} \;.
\end{equation}
Thus $\Omega_H$, $r_{-}$ and $M$  are given in terms of parametric equations $\Omega_H=\Omega_H(r_H,a)$,
$r_{-}= r_{-}(r_H,a)$, and  $M=M(r_H,a)$,  which are provided by Eqs.~(\ref{OmH}), (\ref{r-}) and (\ref{BHmass}), respectively.
These equations will be useful to compute $\Omega_H$, $r_{-}$ and $M$ when finding the values for $a$ that solves the eigenvalue
problem for $\Psi$, given $r_H$.

The so called {\it extremal} Kerr BH corresponds to the following values of the BH properties:
\begin{eqnarray}
a &=& M \;,\\
r_H &=& M=r_{-} \;,\\
\Omega_H&=& \frac{1}{2M}\;,\\
\label{Delta2}
\Delta &=& (r-M)^2 \;.
\end{eqnarray}

Another condition that is imposed on the field $\Psi$ for the boson clouds to exist
is the so-called {\it zero flux} condition at the BH horizon~\cite{Herdeiro2014,Herdeiro2015}\footnote{The no-flux 
condition at the horizon translates into a stationarity ``equilibrium'' condition $\chi^a j_a|_H=0$ where 
$j_a= \frac{1}{\imath}\Big(\Psi^* \nabla_a\Psi- \Psi\nabla_a\Psi^*\Big)$ 
is the conserved current associated with the global phase symmetry of the matter component. This condition ensures that the cloud configuration 
corresponds to {\it bound states}. Furthermore, 
the Killing field ${\tilde \chi}^a := \xi^a + (\omega/m) \eta^a$ is the one which makes 
$\Psi$ to respect also the simultaneous symmetries generated by $\xi^a$ and $\eta^a$: 
${\tilde \chi}^a\nabla_a\Psi\equiv 0$. One then selects ${\tilde \chi}^a\equiv \chi^a$ to coincide with the BH-horizon generator, 
which leads also to the no-flux condition (\ref{fluxcond2})~\cite{Herdeiro2015}. For a boson field with $\omega$ complex 
instead of bound states the configuration may 
undergo {\it superradiance}~\cite{superradiance} if the real part $\omega_R$ lies in the interval $0<\omega_R<m\Omega_H$ 
and the imaginary one $\omega_I>0$, or decay in time when $m\Omega_H<\omega_R$ and $\omega_I<0$ (see \cite{Brito2015} for a review). 
The cloud scenario is precisely the one with $\omega_R=m\Omega_H$ and $\omega_I=0$~\cite{Hod2012,Hod2013,Herdeiro2014,Herdeiro2015}.}:
\begin{eqnarray}
\label{fluxcond}
&& \chi^a \nabla_a \Psi |_{r_H} = 0 \;,\\
\label{chi}
&& \chi^a := \xi^a + \Omega_H \eta^a \;,
\end{eqnarray}
where $\chi^a$ is the {\it helical} Killing vector field given in terms of the {\it timelike} Killing field $\xi^a= (\partial/\partial t)^a$ and the 
{\it axial} Killing field $\eta^a= (\partial/\partial \varphi)^a$. At the horizon $\chi^a$ becomes null and thus, it is tangent to the 
null geodesic generators of the horizon. Equation~(\ref{fluxcond}) together with (\ref{Psians})
lead to the relationship 
\begin{equation}
\label{fluxcond1}
(\omega- m\Omega_H)\Psi_H=0 \;.
\end{equation}
Assuming $\Psi_H\neq 0$ we obtain
\begin{equation}
\label{fluxcond2}
\omega= m\Omega_H \;.
\end{equation}

\section{Obstructions towards a more general no-hair theorem}
\label{sec:obstructions}

Let us now consider Eq.(\ref{KG}). Let $\langle\langle M\rangle\rangle$ denote the domain of outer communication of the BH 
(i.e. $I^{-}(\mathscr{J}^+)\cap I^{+}(\mathscr{J}^-)$) and ${\cal V}\subset \langle\langle M\rangle\rangle$ an open subset bounded by 
the following sets of spacetime points of $M$: two spacelike hypersurfaces 
$\Sigma_1 \subset \langle\langle M\rangle\rangle $, $\Sigma_2 \subset \langle\langle M\rangle\rangle $, a section of the BH horizon 
$H$ (i.e. ${\bar {\cal V}}\cap H$) and, finally by spatial infinity $i^0$. 
We can multiply both sides by $\Psi^*$ and integrate over the spacetime volume ${\cal V}$:
\begin{equation}
\int_{\cal V} \Psi^*\Box \Psi \sqrt{-g}d^4 x = \int_{\cal V} \mu^2 \Psi^* \Psi \sqrt{-g}d^4 x \;.
\end{equation}
Integrating by parts, and using the Gauss theorem one is led to
\begin{equation}
\int_{\partial {\cal V}} \Psi^*  s^c \nabla_c\Psi dS = \int_{\cal V} \Big[ \mu^2 \Psi^* \Psi + (\nabla_c \Psi^*) (\nabla^c\Psi)\Big] \sqrt{-g}d^4 x \;,
\end{equation}
where the surface integral on the left-hand-side (l.h.s) is performed on the boundaries with normal $s^a$: 
the two spacelike hypersurfaces $\Sigma_1$ and $\Sigma_2$, the ``inner'' boundary associated with the subset of the BH horizon ${\bar {\cal V}}\cap H$, and the 
``outer'' boundary at spatial infinity $i^0$. Due to the assumption of stationarity, the integral over the spacelike hypersurfaces 
cancel each other because the integrals are identical except for the opposite sign of their normals. 
On the other hand, at the horizon the normal $s^a$ coincides with the null generator $\chi^a$, and due to the 
zero flux assumption Eq.~(\ref{fluxcond}), the integral vanishes identically, where we assume that $\Psi^*$ is bounded at $H$. 
Finally, the integral at spatial infinity (over a sphere 
with $r\rightarrow \infty$) also vanishes if $\Psi$ falls off sufficiently fast asymptotically (e.g. 
one expects a Yukawa type of fall-off due to the presence of the mass term). Thus one concludes
\begin{equation}
\label{intident1}
\int_{\cal V} \Big[ \mu^2 \Psi^* \Psi + (\nabla_c \Psi^*) (\nabla^c\Psi)\Big] \sqrt{-g}d^4 x \equiv 0 \;.
\end{equation}
In the non-rotating ($\Omega_H\equiv 0\equiv a\equiv \omega)$ and spherically symmetric case ($m=0$) $(\nabla_c \Psi^*) (\nabla^c\Psi)= 
g^{rr}(\partial_r \phi)^2$.
Thus, the previous integral reduces to 
\begin{equation}
\label{intident2}
\int_{\cal V} \Big[ \mu^2 \phi^2 +  g^{rr}(\partial_r \phi)^2 \Big] \sqrt{-g}d^4 x \equiv 0 \;.
\end{equation}
Since $g^{rr}=(1-2M/r) \geq 0$ in the domain of outer communication,
each term is non-negative and thus each one has to vanish independently, leading to $\phi(r)\equiv0$. 
The conclusion is that a nontrivial scalar-field configuration is {\it not} possible within the Schwarzschild spacetime, and thus 
that scalar-clouds are absent. Even if the background is not fixed to be the Schwarzschild spacetime, the
staticity assumption $g^{rr}\geq 0$ in the domain of outer communication together with the asymptotic flatness condition suffices to obtain the same 
conclusion: a static and spherically symmetric spacetime endowed with 
a regular BH region does not allow for a non-trivial hair, and thus, the only possible exterior solution is the Schwarzschild spacetime
\footnote{For a regular BH solution we demand that all the scalars are well behaved in the domain of outer communication, notably, at the horizon $H$. 
In particular, the scalar $(\nabla_c \Psi^*) (\nabla^c\Psi)$ should be well behaved. Other methods to prove the no-hair theorem for complex scalar fields 
were devised in~\cite{Pena1997} using the conservation of the EMT following~\cite{Sudarsky1995}, where the authors assumed in their proof that the scalar $T_{ab}T^{ab}$  
involving the EMT is regular at the horizon, as well as its individual contributions. As they require $T_{r}^{\,\,r}|_H= T_{t}^{\,\,t}|_H$ one is led to the 
conclusion $-g^{tt}\omega^2 \phi^2|_H=0$, which require the condition $\omega^2 \phi^2|_H=0$, given that $g^{tt}$ diverges at $H$.
This condition is similar to the no-flux condition (\ref{fluxcond1}) when $\Omega_H\equiv 0$. From (\ref{Tab}) we appreciate that the components $T_{r}^{\,\,r}$ and $T_{t}^{\,\,t}$ contain the scalar $(\nabla_c \Psi^*) (\nabla^c\Psi)$ and therefore 
the regularity of those components require regularity of this scalar and vice versa. 
When the scalar field is real-valued, i.e., $\omega\equiv 0$, the scalar field is time independent. In this 
case the generator at the horizon is the Killing field $\xi^a$, which becomes null at $H$ and is also the normal of the inner boundary, and thus 
$s^c \nabla_c\Psi=\xi^c  \nabla_c \Psi= \xi^t \partial_t\Psi\equiv 0$. The rest of the proof holds, and we recover the no-hair theorem for the static and spherically 
symmetric BH's with a real-valued scalar field (cf.\cite{Bekenstein1972}).}. This is one of the well known no-hair theorems for a scalar field in the static and spherically symmetric scenario~\cite{Bekenstein1972}. 

Let us consider now a stationary and axisymmetric spacetime, notably the Kerr solution~(\ref{Kerr}),  which is the most relevant for this paper. In this case and for a field with the harmonic dependence (\ref{Psians}) one is led to the following expression for the {\it kinetic} term
\begin{eqnarray}
\label{kinstataxi}
&& K:= (\nabla_c \Psi^*) (\nabla^c\Psi) 
= m^2  \phi^2 \Big[g^{tt}\Omega_H^2 - 2 g^{t\varphi}\Omega_H + g^{\varphi\varphi}\Big] \nonumber \\ 
&+&  g^{IJ}(\nabla_I \phi) (\nabla_J \phi) \;, 
 \end{eqnarray}
where we used Eq.~(\ref{fluxcond2}) in the first line, and the capital indices, which run $1,2$, correspond to the subspace covered by the coordinates 
$r,\theta$. Since in the domain of outer communication of the Kerr BH $g^{rr}\geq 0$ and $g^{\theta\theta}\geq 0$, then
$g^{IJ}(\nabla_I \phi) (\nabla_J \phi)= g^{rr}(\partial_r \phi)^2 + g^{\theta\theta}(\partial_\theta \phi)^2$ is non-negative. The integrand of the volume integral (\ref{intident1}) becomes
\begin{eqnarray}
\label{Integrand}
I &:=& K + \mu^2 \Psi^*\Psi \nonumber \\
&=& m^2 \phi^2 \Big[g^{tt}\Omega_H^2 - 2 g^{t\varphi}\Omega_H  g^{\varphi\varphi} + g^{\varphi\varphi} \Big] \nonumber \\
&+&   g^{IJ}(\nabla_I \phi) (\nabla_J \phi) + \mu^2  \phi^2 \;,
\end{eqnarray}
Thus, if 
non-trivial regular clouds exist (i.e. $\phi(r,\theta)\neq 0$), then {\it a fortiori} the following 
inequality 
\begin{equation}
\label{haircond}
{\cal R}:= g^{tt}\Omega_H^2 - 2 g^{t\varphi}\Omega_H+ g^{\varphi\varphi}\leq 0 \;,
\end{equation}
must hold in a spacetime region in order for the first term of~(\ref{Integrand}) to compensate for the non-negative 
definite terms $g^{IJ}(\nabla_I \phi) (\nabla_J \phi)$ and $\mu^2  \phi^2$ 
and thus for the volume integral (\ref{intident1}) to be satisfied. As we stressed before, in the non-rotating case 
$\Omega_H \equiv 0 \equiv a$, one has $I \geq 0$ and each of the terms of $I$ are also non-negative, thus 
the volume integral (\ref{intident1}) is only satisfied when each of those terms vanish. Then one concludes that 
only the trivial configuration $\phi(r,\theta)\equiv 0$ is possible.

Now, even in the rotating case one would be tempted to prove ${\cal R}\geq 0$, in which case every term of 
$I$ would be positive semidefinite leading again to $\phi(r,\theta)\equiv 0$ for Eq.(\ref{intident1}) to hold. If this were possible then clouds (and more generically, hair) 
would be absent in the rotating case as well and one would have a novel no-hair theorem. However, unless some very restrictive and unphysical 
conditions are assumed, such a task seems impossible. Indeed, the discovery of boson clouds around a extremal and non-extremal 
Kerr backgrounds~\cite{Hod2012,Hod2013,Herdeiro2014,Herdeiro2015,Hod2016} and more generally, 
the existence of boson hair around a stationary, axisymmetric and rotating BH found by Herdeiro and Radu~\cite{Herdeiro2014,Herdeiro2015} provide 
clear evidence that 
an elaboration of a non-hair theorem in the way suggested above seems hopeless. The goal of this paper is then to show that for the 
{\it non-trivial} cloud configurations found in~\cite{Hod2012,Hod2013,Herdeiro2014,Herdeiro2015,Hod2016} the inequality~(\ref{haircond}) holds in several spacetime regions, namely, far from the horizon [cf. Eq.~(\ref{Ibrack}) in Sec.~\ref{sec:eingenprob}]
which allows us to understand in a more heuristic way why Eq.~(\ref{intident1}) is verified when such non-trivial configurations exist, 
while one is usually accustomed that  Eq.~(\ref{intident1}) only holds for trivial scalar-fields in stationary situations. In view of this, we argue that the existence of such clouds 
and hair represents severe obstructions to the extensions of the no-hair theorem 
to the non-spherically symmetric scenarios with complex-valued scalar fields of the sort analyzed here. 

As we will show below, the existence of non-trivial clouds depends on solving an 
eigenvalue problem which ``quantizes'' the possible values for the BH properties, namely, $a$, $\Omega_H$ and $M$, given the {\it quantum numbers} $(n,l,m)$. The {\it principal} number $n$ 
corresponds to the number of nodes associated with the radial part of $\phi(r,\theta)$, while 
the integers $l$ and $m$ are the {\it orbital} and {\it magnetic} quantum numbers, respectively, associated with the angular dependence of the boson field. 
The eigenvalues are represented by the possible values of the BH angular momentum $a$, that can be denoted by $a_{nlm}$, 
which is no longer a continuous parameter but a discrete set labeled by the integers $(n,l,m)$, resulting  from the eigenvalue problem associated with the 
localized solution for $\phi$.

\section{3+1 Splitting, the Klein Gordon Equation and the Kinetic Term at the Horizon}
\label{sec:eingenprob}

The d'Alambertian operator of Eq.(\ref{KG}) with the Kerr background metric (\ref{Kerr}) reads
\begin{eqnarray}
&& \Box \Psi = \frac{1}{\sqrt{-g}}\partial_a \Big[\sqrt{-g} g^{ab}\partial_b\Psi\Big] \nonumber \\
&=& D^2_{r\theta} \Psi - \Big[g^{tt}\omega^2 - 2m g^{t\varphi}\omega+ m^2 g^{\varphi\varphi}\Big]\Psi \;, \\
&& \!\!\!\!\!\!\!\!\!\! D^2_{r\theta}\Psi := \frac{1}{\sqrt{-g}}\partial_I \Big[\sqrt{-g} g^{IJ}\partial_J\Psi\Big] \nonumber \\
&&\!\!\!\!\!\!\!\!\!\!= \frac{1}{\sqrt{-g}}\left(\partial_r \Big[\sqrt{-g} g^{rr}\partial_r\Psi\Big] 
+ \partial_\theta \Big[\sqrt{-g} g^{\theta\theta}\partial_\theta\Psi\Big]\right)\;.
\end{eqnarray}
These results together with Eqs.~(\ref{Psians}) and (\ref{fluxcond2}), allow us to write the Klein-Gordon Eq.~(\ref{KG}) 
as an eigenvalue problem for the amplitude $\phi(r, \theta)$ in the form
\begin{equation}
\label{eigen1}
D^2_{r\theta}\phi = \Big[m^2(g^{tt}\Omega_H^2 - 2 g^{t\varphi}\Omega_H +  g^{\varphi\varphi}) +\mu^2 \Big] \phi \;,
\end{equation}
where the eigenvalues, represented by the possible values of the angular momentum $a$, are {\it hidden} within the expression in brackets. 
In Section~\ref{sec:numerics} below we present more explicitly the eigenvalue problem at hand. From the r.h.s of Eq.~(\ref{eigen1}) we appreciate that the terms within the brackets not involving the mass term $\mu^2$ are exactly the same that appear in Eq.~(\ref{kinstataxi}). Of course, this is not surprising taking into account that Eq.~(\ref{intident1}) is the result of integrating Eq.~(\ref{eigen1}).
Thus, the existence of non-trivial clouds is related to the 
existence of a non-trivial solution to the eigenvalue problem for $\phi(r,\theta)$ which in turn is closely related with the 
verification of the inequality (\ref{haircond}).

It is useful to write Eq.~(\ref{kinstataxi}) in terms of the 3+1 variables~\cite{3+1} as follows\footnote{Indices $i,j$ run $1-3$.}
\begin{eqnarray}
\label{kinstataxi2}
&& K=  m^2 \phi^2 {\cal R} +  h^{IJ} (D_I\phi) (D_J\phi) \;, \\
\label{haircond2}
&& {\cal R} = - \Big[ \left(\frac{\Omega_H - N^{\varphi}}{N}\right)^2 - h^{\varphi\varphi}\Big] \;,\\
&& g_{ij}= h_{ij} \;,\\
&&  g_{tt}= - N^2 + h_{ij}N^i N^j \;,\\
&&  g_{ti}= - N_i= - h_{ij}N^i \;, \\
&& g^{ij}= h^{ij}- \frac{N^i N^j}{N^2} \;,\\
&& g^{tt}= -\frac{1}{N^2} \;,\; g^{it}= -\frac{N^i}{N^2} \;,\\
&& g^{t\varphi} = -\frac{N^{\varphi}}{N^2} \;,\\
&& g^{\varphi\varphi}= h^{\varphi\varphi} -\frac{(N^{\varphi})^2}{N^2} \;,\\
&& h^{\varphi\varphi}= \frac{\rho^2}{[(r^2+a^2)^2-\Delta a^2\sin^2\theta]\sin^2\theta} \;,\\
\label{g^rr}
&& g^{rr}= h^{rr} -\frac{(N^{r})^2}{N^2}= h^{rr} =\frac{1}{h_{rr}}= \frac{\Delta}{\rho^2}\;,\\
\label{g^tettet}
&& g^{\theta\theta}= h^{\theta\theta} -\frac{(N^{\theta})^2}{N^2}=  h^{\theta\theta} =\frac{1}{h_{\theta\theta}}= \frac{1}{\rho^2}\;,\\
&& h^{IJ} (D_I\phi) (D_J\phi) = h^{rr}(\partial_r\phi)^2 + h^{\theta\theta}(\partial_\theta\phi)^2\;,
\end{eqnarray}
where we used in Eqs.~(\ref{g^rr}) and (\ref{g^tettet}) the fact that in the axisymmetric problem at hand the shift vector components 
$N^r$ and $N^\theta$ vanish identically.
Furthermore, $g_{rr}= h_{rr}$, $g_{\theta\theta}=h_{\theta\theta}$, $g_{\varphi\varphi}=h_{\varphi\varphi}=1/h^{\varphi\varphi}$. 
The covariant derivatives $D_I$ are associated with the 3-metric $h_{ij}$, and $h^{ij}$ is the 
inverse of the 3-metric. One can show that the lapse function is given by~\cite{Shibata2016}
\begin{equation}
N^2= \frac{\Delta \rho^2}{(r^2+a^2)^2-\Delta a^2\sin^2\theta} \;,
\end{equation}
and the square of the helical Killing vector field (\ref{chi}) at the horizon can be expressed in terms of the lapse function 
(cf. footnote~\ref{foot:BH3+1}): 
\begin{eqnarray}
\chi^a \chi_a |_H &=& \Big[\xi^a \xi_a + 2\Omega_H \xi_a \eta^a + \Omega_H^2 \eta_a \eta^a \Big]_H\nonumber \\
              &=& \Big[g_{tt} + 2\Omega_H g_{t\varphi} + \Omega_H^2 g_{\varphi\varphi}  \Big]_H= -N^2_H\;.
\end{eqnarray}
Thus, the horizon is located at the place where $N(r,\theta)=0$ and also where $g_{rr}=\infty$, i.e., where $\Delta=0$ 
[cf. Eqs.~(\ref{Kerr}) and (\ref{Delta})]. From Eqs.~(\ref{kinstataxi2}) and (\ref{haircond2}) we notice that regularity
of the kinetic term $K$ at the horizon requires $N^\varphi_H=\Omega_H$ and $h^{\varphi\varphi}_H<\infty$, 
as well as $h^{IJ} (D_I\Psi^*) (D_j\Psi)|_H= [h^{rr}(\partial_r\phi)^2+ h^{\theta\theta}(\partial_\theta\phi)^2]_H<\infty$. 
Since $h^{rr}|_{r_H}\equiv 0$ and $h^{\theta\theta}|_{r_H}<\infty$ [cf. Eqs.~(\ref{g^rr}) and (\ref{g^tettet})], 
it is required that $\partial_r\phi$ and $\partial_\theta\phi$ 
be bounded at the horizon or at least that $h^{rr}(\partial_r\phi)^2|_{r_H}<\infty$, and $h^{\theta\theta}(\partial_\theta\phi)^2|_{r_H}<\infty$. 
In the next section we will analyze the regularity conditions for the field and its radial derivatives 
at the horizon using the Teukolsky equation and elaborate more about the regularity of the kinetic term at the horizon, notably, 
in the (near) extremal case. 

Now, as concerns the condition $N^{\varphi}_H=\Omega_H$, this is precisely the definition of the 
rotating frequency of the horizon. The term
\begin{equation}
\left(\frac{\Omega_H - N^{\varphi}}{N}\right)^2 \Big|_H \;,
\end{equation}
leads thus to $0/0$ and it requires a careful analysis in order to evaluate its precise value, specially in the extremal case. Perhaps the easiest way to compute such a value is to expand the shift component 
$N^{\varphi}(r,\theta)$ in Taylor series near the horizon and use the fact that $N^2$ can be written in terms of the product of two factors involving 
the two horizons $r_H$ and $r_{-}$:
\begin{eqnarray}
&& N^{\varphi}(r,\theta)= \frac{2aMr}{(r^2+a^2)^2 -\Delta a^2 \sin^2\theta}  \;,\\
&& N^{\varphi}(r,\theta)= N^{\varphi}(r_H,\theta)+ \partial_rN^{\varphi}(r,\theta)|_{r=r_H} (r-r_H) 
\nonumber \\
&& \;\;\;\;\;\;\;\;\;\;\;\;\;\;\;\;\;\; + {\cal O}((r-r_H)^2) \;,\\
&& N^{\varphi}(r_H,\theta) = \Omega_H \;,\\
&& \partial_rN^{\varphi}(r,\theta)|_{r=r_H} = \frac{a}{4M^3 r_H^3}\Big[2Mr_H(M-2r_H) \nonumber \\
&&  \;\;\;\;\;\;\;\;\;\;\;\;\;\;\;\;\;\;\;\;\;\;\;\;\;\;\;\;  + a^2(r_H-M)\sin^2\theta\Big] \;,\\
&& N^2 = \Delta \sin^2\theta h^{\varphi\varphi}= (r-r_H)(r-r_{-}) \sin^2\theta h^{\varphi\varphi} \;.
\end{eqnarray}
Thus, for the non-extremal and the extremal scenarios we find
\begin{eqnarray}
\label{ommNphi}
&&\left(\frac{\Omega_H - N^{\varphi}}{N}\right)^2 \Big|_H =0 \;\;\;\;\;\;\;\;\;\;\;\;(r_{-} \neq r_H)\;,\\
\label{ommNphiext}
&&\left(\frac{\Omega_H - N^{\varphi}}{N}\right)^2 \Big|_H^{\rm ext} = 
\frac{\Big[\partial_rN^{\varphi}(r,\theta)\Big]^2}{\sin^2\theta h^{\varphi\varphi}} |_{r=r_H=M}\nonumber  \\
&& = \frac{1}{M^2(1+\cos^2\theta)} \;\;\;\;\;\;\;\;\;\;\;\;\;\;(r_{-} = r_H=M=a)\;,
\end{eqnarray}
where we used
\begin{eqnarray}
h^{\varphi\varphi}|_H &=& \frac{r_H^2 + a^2\cos^2\theta}{4M^2 r_H^2 \sin^2\theta} \;,\\
 h^{\varphi\varphi}|^{\rm ext}_H &=& \frac{1 + \cos^2\theta}{4M^2 \sin^2\theta} \;.
\end{eqnarray}
A straightforward, albeit longer computations, allow to verify that both (\ref{ommNphi}) and (\ref{ommNphiext}) also hold when the exact 
Kerr metric is used instead of expanding the shift component in Taylor series near the horizon.

From the above results and from Eq.~(\ref{haircond2}) we conclude
\begin{eqnarray}
{\cal R}_H &=& h^{\varphi\varphi}|_H  \;, \\
{\cal R}_H^{\rm ext} &=& \frac{4\sin^2\theta - (1+\cos^2\theta)^2}{4M^2(1-\cos^4\theta)} \;.
\end{eqnarray}
In particular at $\theta=\pi/2$,
\begin{eqnarray}
{\cal R}_H &=& \frac{1}{4M^2} \;, \\
{\cal R}_H^{\rm ext} &=& \frac{3}{4M^2} \;,
\end{eqnarray}
which are both positive.

\bigskip
On the other hand, far from the horizon, we have the following asymptotic behavior
\begin{eqnarray}
N^2 &\sim& 1 + {\cal O}(1/r) = 1 - \frac{2M}{r}\;,\\
N^{\varphi} &\sim& {\cal O}(1/r^3) = \frac{2aM}{r^3}\;,\\
h^{\varphi\varphi} &\sim& {\cal O}(1/r^2)= \frac{1}{r^2 \sin^2\theta}\;.
\end{eqnarray}
So asymptotically,
\begin{eqnarray}
&& \left(\frac{\Omega_H - N^{\varphi}}{N}\right)^2 - h^{\varphi\varphi} \sim \Omega_H^2\left(1+\frac{2M}{r}\right) 
+\frac{1}{r^2 \sin^2\theta}\nonumber \\
&& \sim  \Omega_H^2 -\frac{1}{r^2 \sin^2\theta}\;,
\end{eqnarray}
From this result we conclude  that asymptotically Eq.~(\ref{haircond2}) behaves
\begin{equation}
{\cal R} \sim -\Omega_H^2 +\frac{1}{r^2 \sin^2\theta}  \;.
\end{equation}
This result is valid for the extremal and non-extremal scenarios.

Far from the axis of symmetry ($\theta=0,\pi$), in particular at the equatorial plane $\theta=\pi/2$,
and for a finite rotation frequency, the condition (\ref{haircond}) holds asymptotically
\begin{equation}
{\cal R} \sim  -\Omega_H^2 <0\;.
\end{equation}
Therefore asymptotically and at $\theta=\pi/2$  the first term of Eq.(\ref{kinstataxi2}) behaves
\begin{eqnarray}
\label{Ibrack}
 \Lambda:= m^2 \phi^2 {\cal R} &=& - m^2 \phi^2\Big[ \left(\frac{\Omega_H - N^{\varphi}}{N}\right)^2 - h^{\varphi\varphi}\Big] \nonumber \\
&\sim& -m^2\phi^2 \Omega_H^2 <0\;.
\end{eqnarray}
Clearly when $\Omega_H\equiv 0$, we have $\Lambda= m^2\phi^2 h^{\varphi\varphi}$, which is positive semidefinite. 
In particular, at $\theta=\pi/2$,  $\Lambda=m^2\phi^2/r^2$. In this case one finds again that the kinetic term Eq.~(\ref{kinstataxi2})
is positive semidefinite and we require $\phi(r,\theta)\equiv 0$ for Eq.~(\ref{intident1}) to be satisfied.

In summary, at $\theta=\pi/2$
\begin{eqnarray}
\Lambda|_H &=& m^2 \phi_H^2 {\cal R}_H=  \frac{m^2 \phi_H^2}{4M^2} \;,\\
\Lambda|_H^{\rm ext} &=& m^2 {\phi_H^{\rm ext}}^2 {\cal R}_H^{\rm ext}=  \frac{3m^2 {\phi_H^{\rm ext}}^2}{4M^2} \;,\\
\label{Lambdaasym}
\Lambda &\sim& - m^2 \phi^2 \Omega_H^2 \;\;\;\;,\;\;\;\;(r_H\ll r) 
\end{eqnarray}
where, as emphasized above, the asymptotic behavior for $\Lambda$ holds for the non-extremal and the extremal cases as well. 
We conclude that the rotating contribution to the kinetic term (\ref{kinstataxi2}) is positive at the horizon and then becomes negative. Since 
$\phi$ will vanish asymptotically then $\Lambda$, and in fact all the kinetic term, will vanish asymptotically.

We have reached this conclusion even without solving the eigenvalue problem for $\phi(r,\theta)$. So from this result we have clear  
indications that if a non trivial localized solution for $\phi(r,\theta)$ exists, the inequality (\ref{haircond}) actually holds in some region 
of the outer communication of the BH. 
In the next section we solve the eigenvalue problem numerically and, for simplicity, plot the quantity $\Lambda/\phi^2= m^2  {\cal R}$ at $\theta=\pi/2$
from the horizon $r=r_H$ to $r\gg r_H$ in order to appreciate its negative behavior in a large portion of the spacetime not only asymptotically 
(cf. Figure~\ref{fig:kinetic}), showing that the inequality (\ref{haircond}) actually holds in some regions of spacetime when clouds exist.

\section{The Teukolsky Equation, Regularity Conditions and Numerical results}
\label{sec:numerics}


In order to solve the Klein-Gordon Eq.~(\ref{KG}) or its reduced form Eq.(\ref{eigen1}) the complex-valued scalar field (\ref{Psians}) is 
further decomposed as\footnote{Here we introduced explicitly the subindices in the field referring to the ``quantum'' numbers $n,l,m$.}
\begin{equation}
\Psi_{nlm}\left(t, r, \theta, \phi\right) = R_{nlm}\left(r\right)S_{lm}\left(\theta\right)e^{im\phi}e^{-i\omega t},
\end{equation}
The angular functions $S_{lm}\left(\theta\right)$ are the spheroidal harmonics which obey the angular equation

\begin{widetext}
\begin{equation}
 \label{angularE}
\frac{1}{\sin(\theta)}\frac{d}{d\theta}\left(\sin\theta\frac{dS_{lm}}{d\theta}\right) + \left(K_{lm} + a^2(\mu^2 - 
\omega^2)\sin^2(\theta) - \frac{m^2}{\sin^2(\theta)}\right)S_{lm} = 0 \;.
\end{equation}
\end{widetext}
where $K_{lm}$ are separation constants. We consider the following expansion for the coupling constant $K_{lm}$
\begin{equation}
\label{constan K}
K_{lm} + a^2(\mu^2 - \omega^2) = l(l + 1) + \sum_{k=1}^{\infty}c_{k}a^{2k}(\mu^2 - \omega^2)^k, 
\end{equation}
where the expansion coefficients $c_k$ are given in Ref.~\cite{Abramowitz}.

The radial functions $R_{nml}(r)$ obey the radial Teukolsky equation \cite{Teukolsky}
\begin{widetext}
\begin{equation}
\label{radialE}
\Delta\frac{d}{dr}\left(\Delta\frac{dR_{nlm}}{dr}\right) + \left[{\cal H}^2 + \left(2ma\omega - K_{lm} - \mu^2\left(r^2 + a^2\right)\right)\Delta\right]R_{nlm} = 0  
\end{equation}
\end{widetext}
where 
\begin{equation}
\label{H}
{\cal H}:=\left(r^2 + a^2\right)\omega - am = \frac{ma}{2M r_H}(r-r_H)(r+r_H) \;.
\end{equation}
The last equality arises from the condition (\ref{fluxcond2}) and from Eqs.~(\ref{OmH}) and (\ref{BHmass}). Thus we appreciate that 
${\cal H}$ vanishes at $r=r_H$.

As mentioned in Section~\ref{sec:clouds} 
we are interested in finding {\it bound states} of the field configuration (i.e. boson clouds) corresponding to frequencies $\omega$ satisfying 
Eq.~(\ref{fluxcond2}) and these scalar configurations are characterized by three ``quantum numbers'' $(n, l, m)$. 
The numbers $n,l$ are non-negative integers, $n$ determines the number of nodes of the radial function $R_{nlm}$, 
and the ``magnetic'' number $m$ is an integer satisfying $|m| \leq l$.
In the {\it non-extremal} scenario $(0 < a < M)$,\footnote{We focus only on positive values  for $a$.} for a given horizon radius $r_H$ and for a fixed $l, m$ there is a specific value for 
$a$, i.e., $a_{nlm}$ that leads to a localized configuration, i.e., one where the radial function vanishes asymptotically 
with $n$ nodes. In turn, the mass $M$ and the angular velocity $\Omega_H$ of the BH are also ``quantized'' from Eqs.(\ref{OmH}) and (\ref{BHmass}). 

In order to find {\it genuine} cloud configurations one has to impose regularity conditions of the field $\Psi$ at the BH horizon. 
For instance, the field and its derivatives must be bounded at the horizon. This ensures that several {\it scalars} computed from such 
derivatives are also regular (i.e. bounded) there, namely, the kinetic term (\ref{kinstataxi}). On the other hand, if one assumes that 
the field is $C^3$ at the horizon, the radial derivatives must be finite there as well. In order to find regular solutions of 
Eq.(\ref{radialE}) one requires $R''_{nlm}(r_H)$, $R'_{nlm}(r_H)$ and $R_{nlm}(r_H)$, where {\it primes} indicate radial derivatives. Assuming that $R''_{nlm}(r_H)$ is bounded in Eq.(\ref{radialE}), the regularity condition for $R'_{nlm}(r_H)$ found from Eq.(\ref{radialE}) for the {\it non-extremal} case ($0<a<M$) turns out to be
\begin{equation}
\label{CNE1}
R_{nlm}'(r_H) = -\frac{2m^2a\Omega_{H} - K_{lm} - \mu^2(r_H^2 + a^2)}{2(r_{H} - M)}R_{nlm}(r_{H}) \;.
\end{equation}

The value $R_{nlm}(r_{H})$ is not constrained and we choose $R_{nlm}(r_{H})=1$ for simplicity and in order to compare our numerical results 
with previous studies~\cite{Herdeiro2014,Herdeiro2015,Benone2014}.
 
To find $R''_{nlm}(r_H)$ we need to differentiate Eq.(\ref{radialE}) one more time and demand that $R'''_{nlm}(r_H)$ is 
bounded. We find
\begin{widetext}
\begin{eqnarray}
\label{CNE2} 
\resizebox{1\hsize}{!}{$
\nonumber R_{nlm}''(r_{H})  =  -\frac{1}{4(r_{H} - M)}\left\{\Big[2\left(1 + m^2a\Omega_{H}\right) - K_{lm}
 - \mu^2\left(r_{H}^2 + a^2\right)\Big]R_{nlm}'(r_{H}) 
+ \left( \frac{2m^2a^2(M - r_{-})}{M^2\left(r_H - r_{-}\right)^2} - 2\mu^2r_{H}\right)R_{nlm}(r_{H})\right\}\;,$} \nonumber\\
\end{eqnarray}
\end{widetext}

Notice that in the extremal case $r_H=M=a=r_{-}$ both regularity conditions (\ref{CNE1}) and (\ref{CNE2}) blow up at the horizon. 
Thus, from those regularity conditions we can approach the extremal solutions from the non-extremal ones only in the limit 
$r_H\rightarrow M$. We have done so and we elaborate more about those specific solutions below. In a future investigation we shall analyze if extremal solutions 
with bounded derivatives at the horizon are possible.
 
In order to find boson-cloud solutions embedded in a non-extremal Kerr BH, we solved Eq.~(\ref{radialE}) numerically with their respective regularity conditions (\ref{CNE1}) and (\ref{CNE2}) by integrating in the domain of outer communication. We integrate the Teukolsky Eq.~(\ref{radialE}) from $r=r_H$ outwards using a 4th order 
Runge-Kutta scheme and a {\it shooting} method to find the correct eigenvalues (the angular momentum $a$) to a very good precision. The 
correct eigenvalue is such that the radial function vanishes asymptotically.

\begin{figure}[h]
  \centering
    \includegraphics[width=0.5\textwidth]{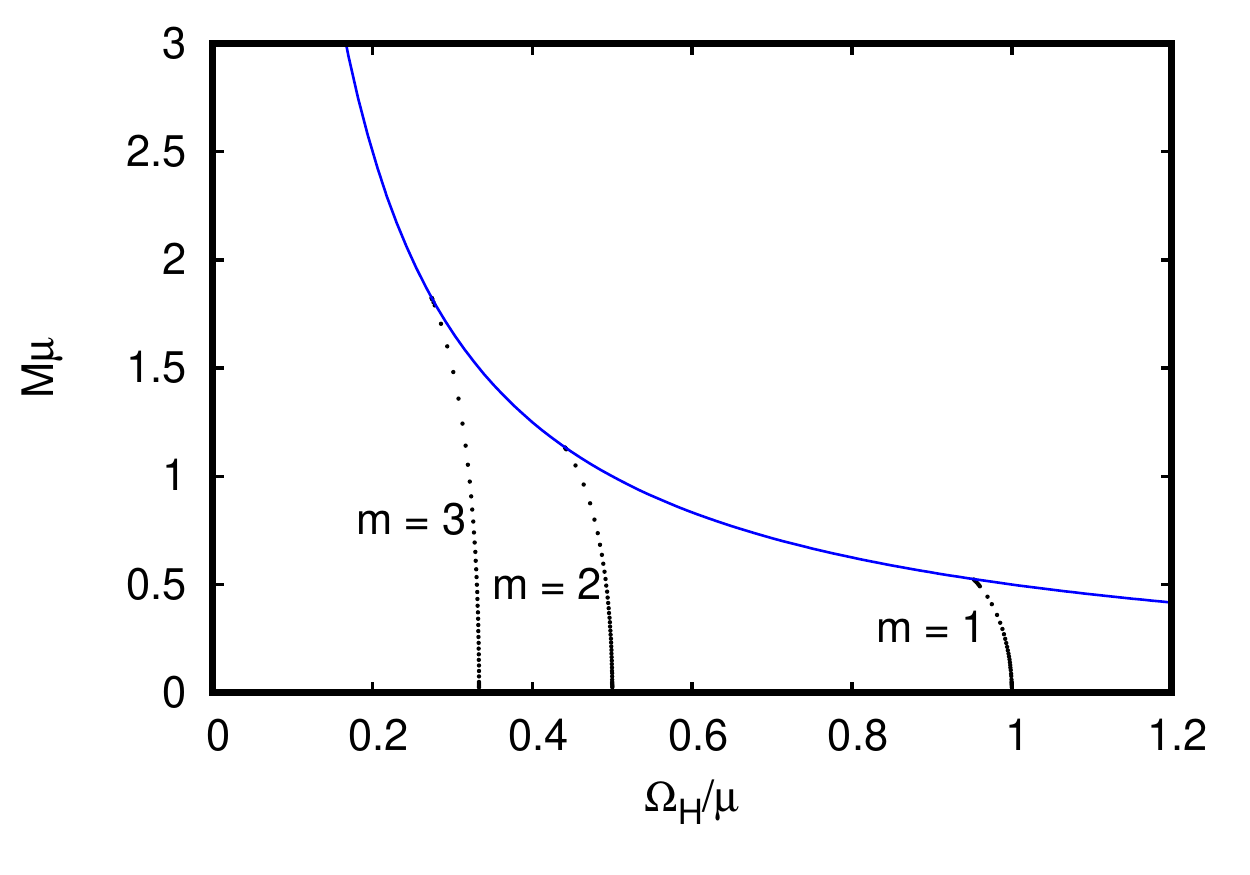}
  \caption{(color online) The dotted lines represent the ``quantized'' (i.e. discrete) values for 
the mass $M$ and angular velocity $\Omega_{H}$ 
(in units of $1/\mu$ and $\mu$ respectively) of the Kerr metric that allow for the existence of boson clouds in the non-extremal case. These values 
are found from the eigenvalues $a_{nlm}$ (see Tables~\ref{tab:spectrum1}-\ref{tab:spectrum3}) associated with the fundamental mode $n=0$ and 
$l=m$ with $m=1,2,3$ leading to a localized solution for the radial function $R_{nlm}$ (see Figure~\ref{fig:R11}). 
The blue solid curve represents the extremal case $M= 1/(2\Omega_{H})=a$ (Kerr solutions do not exist above this line). The black dots 
very near the blue curve represent the specific values of $M, \Omega_H$ for which boson clouds exist in the {\it near extremal} situations.}
  \label{fig:MvsOm}
\end{figure}

\begin{figure}[h]
  \centering
\includegraphics[width=0.5\textwidth]{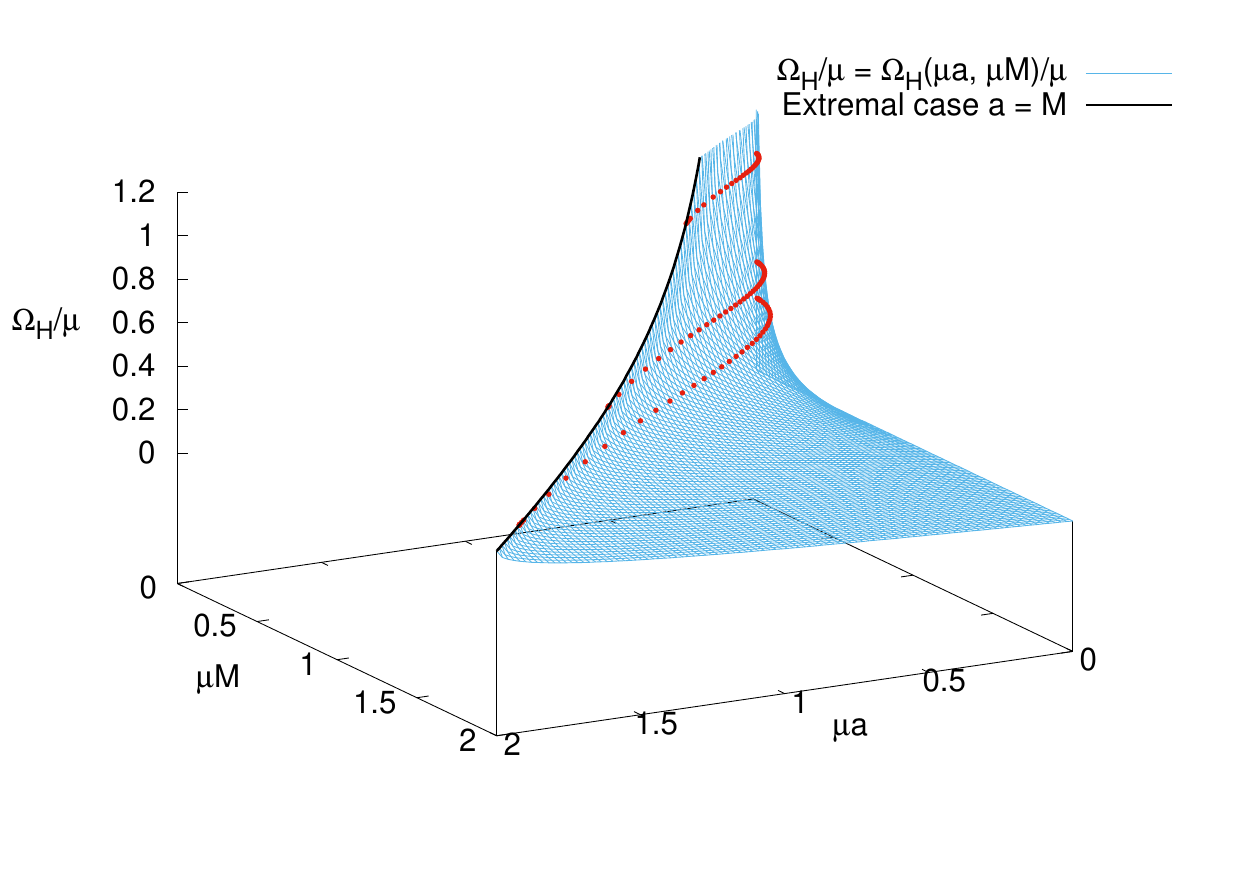}
  \caption{(color online) Rotation frequency $\Omega_H$ as a function of $M$ and $a$ [cf. Eq.~(\ref{OmH})]. The black line corresponds to the  
    the extremal Kerr BH (depicted in blue in Figure~\ref{fig:MvsOm}). Superposed are the values of $a,M,\Omega_H$ (indicated by red dots) associated with the existence of 
clouds. The same values for $M$ and $\Omega_H$ are also depicted in the 2-dimensional plot of Fig.\ref{fig:MvsOm} (black dots).}
  \label{fig:OmvsMa}
\end{figure}

\begin{figure}[h]
  \centering
    \includegraphics[width=0.5\textwidth]{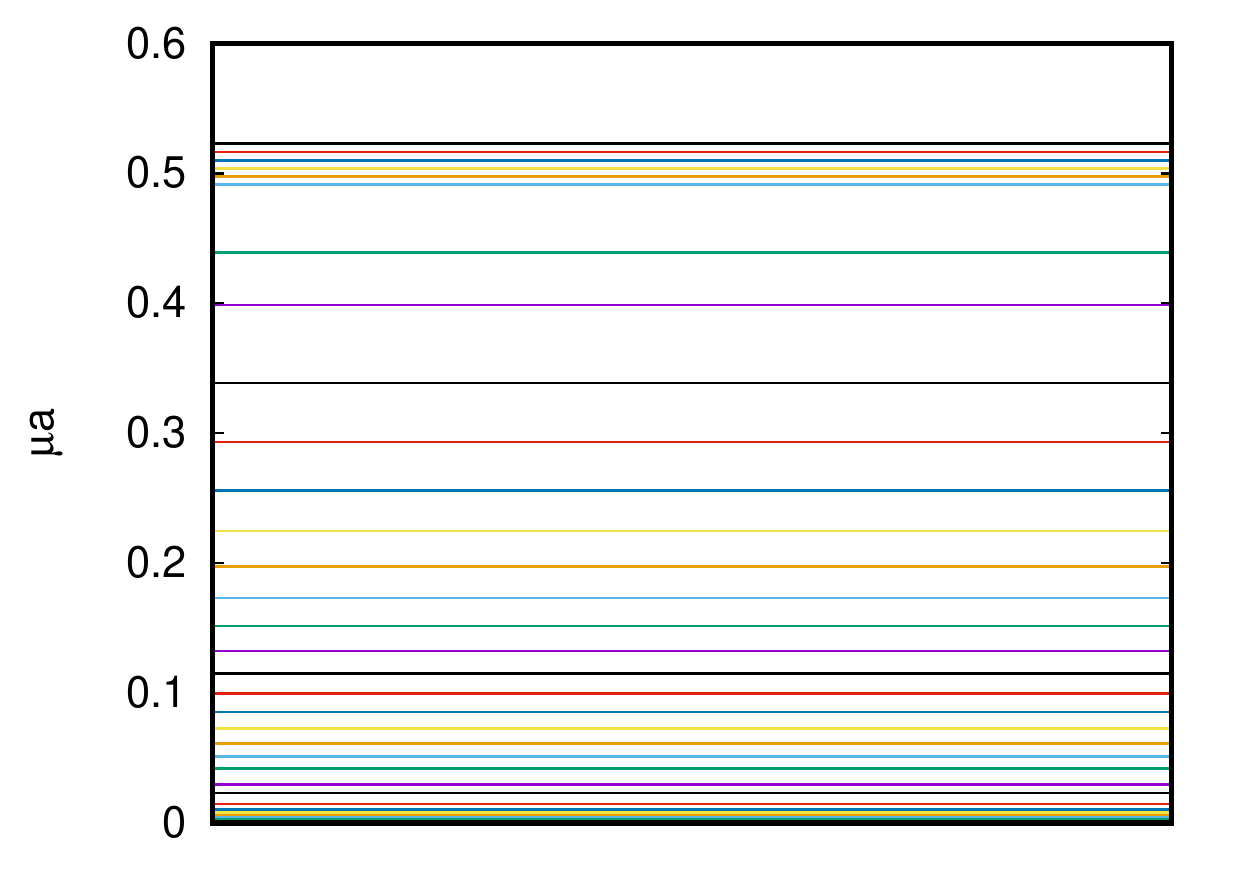}
  \caption{Eigenvalues $a$ (in units $1/\mu$) of the {\it fundamental mode} with numbers $n = 0$ (zero nodes) and $l = m = 1$ for different 
locations of the horizon $r_H$ (not depicted). The precise eigenvalues used in this plot and the corresponding $r_H$ are provided in 
Table~\ref{tab:spectrum1}. These eigenvalues are associated with the dotted line marked $m=1$ in Figure~\ref{fig:MvsOm}.  }
  \label{fig:spectrum1}
\end{figure}

\begin{figure}[h]
  \centering
    \includegraphics[width=0.5\textwidth]{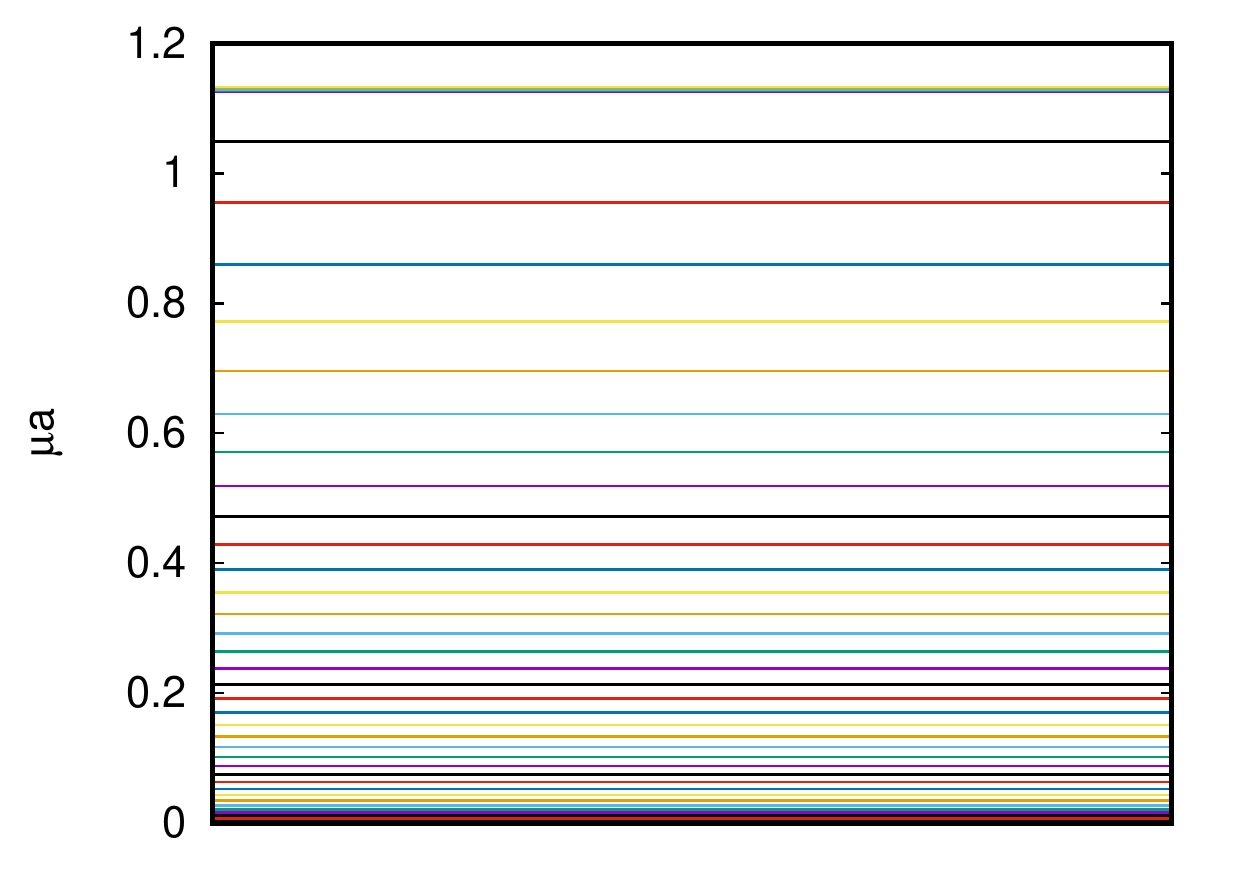}
  \caption{Similar to Figure~\ref{fig:spectrum1} but for the mode $l = m = 2$ with $n = 0$. See Table~\ref{tab:spectrum1}. 
These eigenvalues are associated with the dotted line marked $m=2$ in Figure~\ref{fig:MvsOm}. }
  \label{fig:spectrum2}
\end{figure}

\begin{figure}[h]
  \centering
    \includegraphics[width=0.5\textwidth]{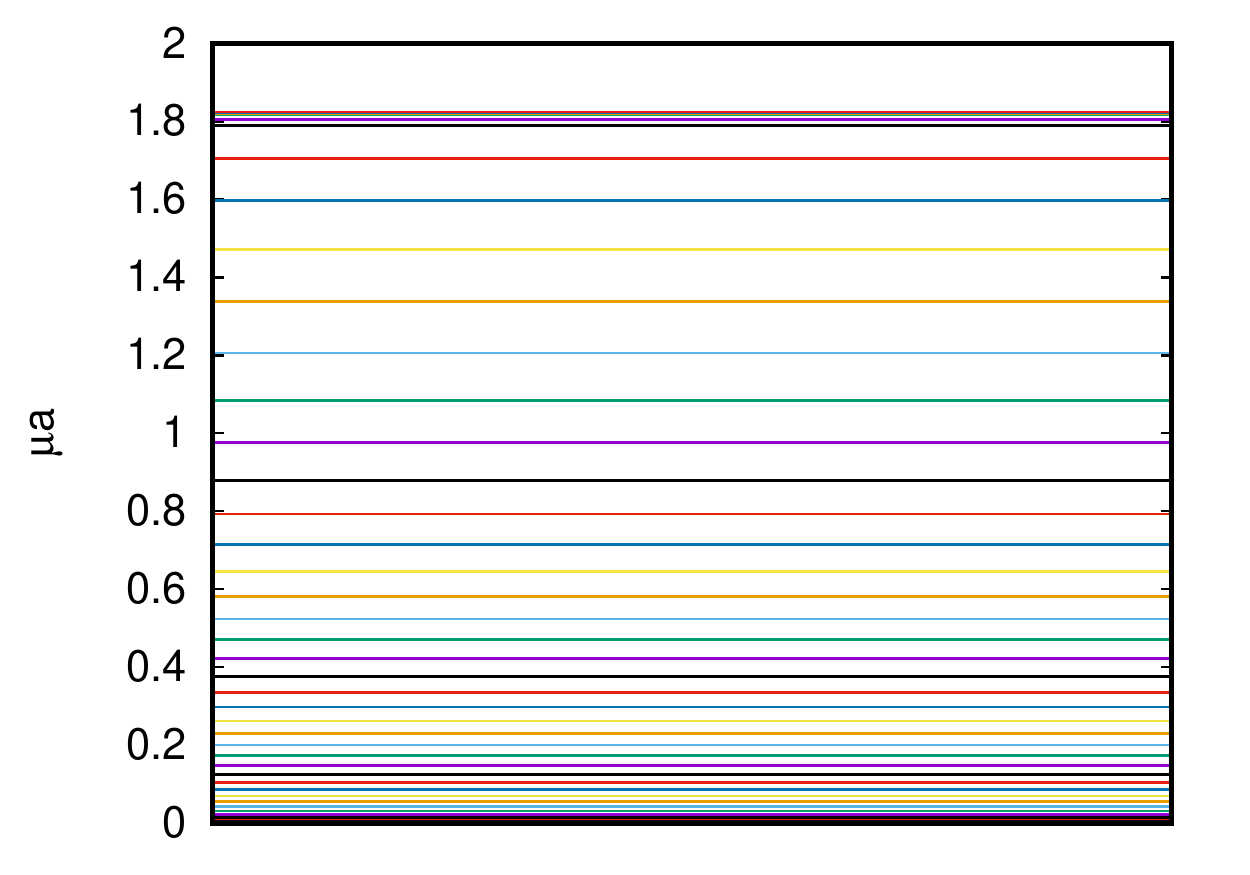}
\caption{Similar to Figure~\ref{fig:spectrum1} for the mode $l = m = 3$ with $n = 0$. See Table~\ref{tab:spectrum3}. 
These eigenvalues are associated with the dotted line marked $m=3$ in Figure~\ref{fig:MvsOm}. }
  \label{fig:spectrum3}
\end{figure}

Figures~\ref{fig:MvsOm} and \ref{fig:OmvsMa} depict the ({\it existence lines}) 
found from Eqs.(\ref{OmH}) and (\ref{BHmass}) when the eigenvalue problem Eq.(\ref{radialE}) is solved numerically. As we stressed above, 
the angular momenta $a_{nlm}$ are the {\it eigenvalues} of the problem (i.e. the spectra). These {\it spectra} and their precise numerical values
are depicted and displayed, respectively, by Figures~\ref{fig:spectrum1}-\ref{fig:spectrum3} and Tables~\ref{tab:spectrum1}-\ref{tab:spectrum3} 
at the end of the paper. Those figures and tables include the near extremal solutions where $M \approx a \approx r_H$. 
Our numerical results are compatible with \cite{Herdeiro2014,Herdeiro2015,Benone2014}.

\begin{figure}[h]
  \centering
    \includegraphics[width=0.5\textwidth]{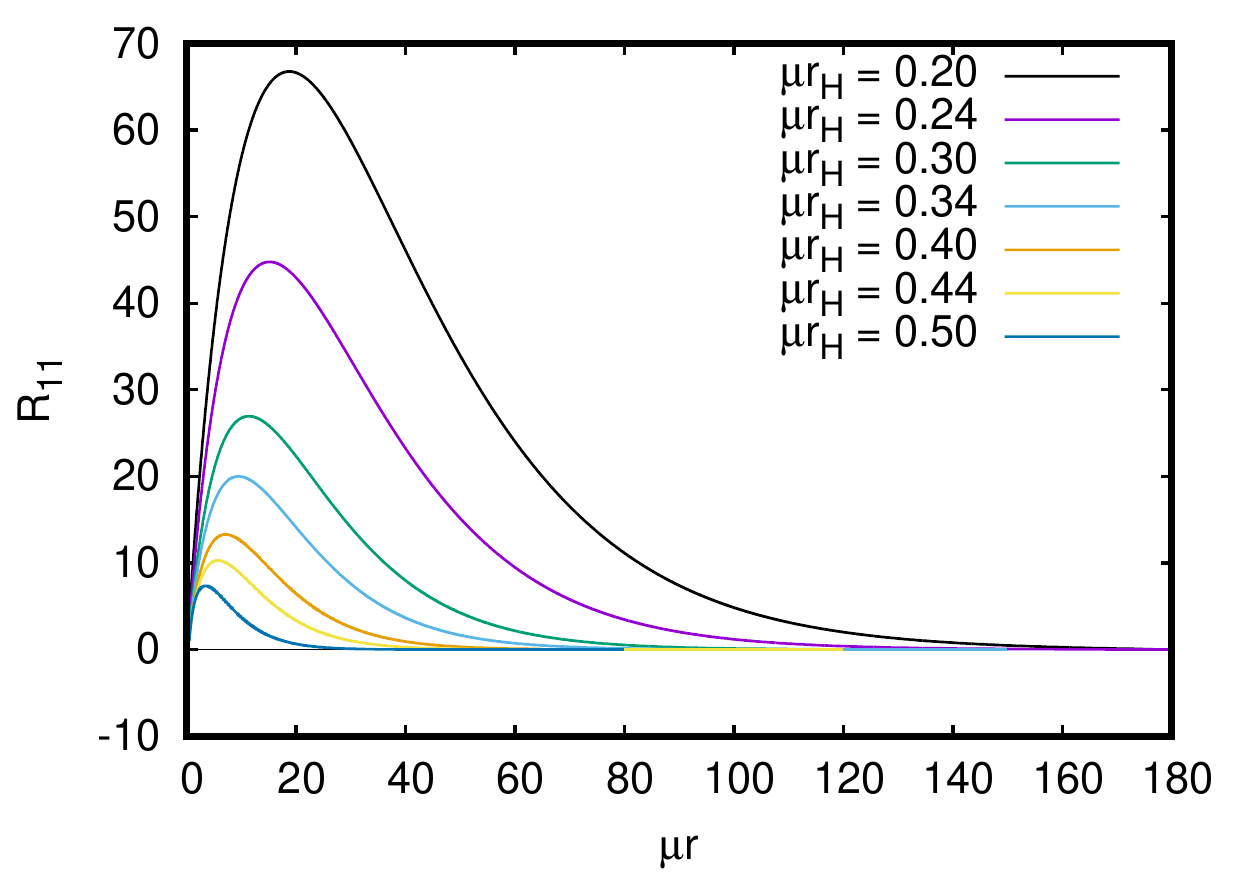}
  \caption{(color online) Sample of radial solutions $R_{11}$ ($R_{011}$) with $n = 0$ and $l=1=m$ associated with boson clouds in Kerr backgrounds with 
different horizon locations $\mu r_{H}$ (non-extremal case). Their corresponding eigenvalues $a_{011}$ are included in Figure~\ref{fig:spectrum1} 
and listed in Table~\ref{tab:spectrum1}, while their values $M$ and $\Omega_H$ correspond to some of the (black) dots marked 
$m=1$ in Figure~\ref{fig:MvsOm} (see also  Figure~\ref{fig:Ranearext}). }
  \label{fig:R11}
\end{figure}

\begin{figure}[h]
  \centering
    \includegraphics[width=0.5\textwidth]{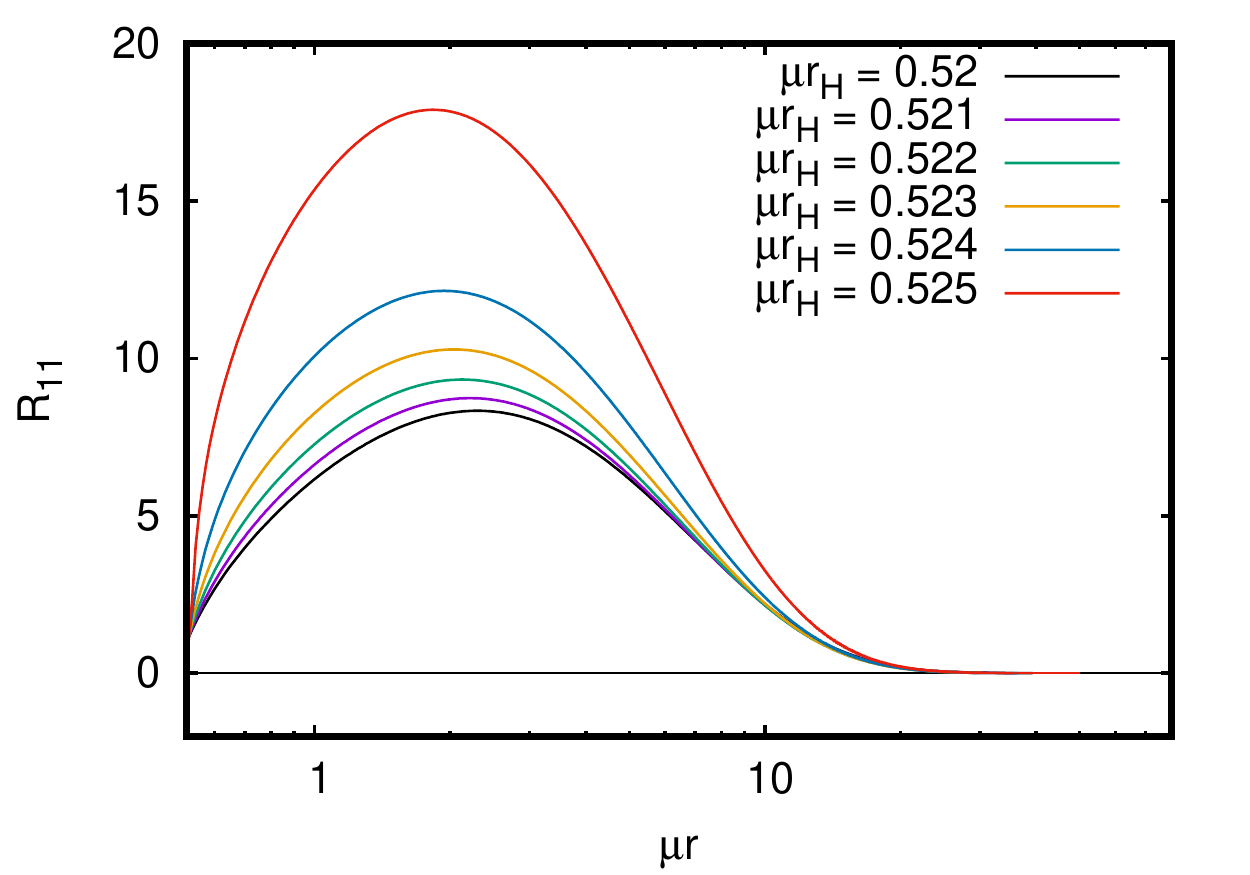}
  \caption{(color online) Similar to Figure~\ref{fig:R11} ($n = 0, l=1=m$) but for 
the near extremal case $r_H\approx M\approx a$. Their corresponding eigenvalues $a_{011}$ are included in Figure~\ref{fig:spectrum1} 
and listed in Table~\ref{tab:spectrum1}, while their values $M$ and $\Omega_H$ correspond to the (black) dots marked
$m=1$ in Figure~\ref{fig:MvsOm} that are close to the (blue) extremal curve.}
  \label{fig:Ranearext}
\end{figure}

Figure~\ref{fig:R11} shows a sample of radial solutions for $n = 0$ and $l=1=m$ using different horizon sizes 
for the non-extremal case. Those solutions have no nodes and vanish asymptotically. Figure~\ref{fig:Ranearext} is similar to 
Fig.~\ref{fig:R11} but for the near extremal case, however unlike  Figure~\ref{fig:R11}, as the configurations approach the extremal case the maximum amplitude 
of the radial function increases as well as the slope at the horizon instead of decreasing.

We emphasize that extremal solutions with unbounded values for $R_{nlm}'(r_H)$ may make sense physically. 
From Eq.~(\ref{CNE1}) we notice that $R_{nlm}'$ blows up at the extremal horizon $r_H=M= a$ as 
$R_{nlm}'= C/(r_H-M)$, where $C$ stands for the numerator of Eq.~(\ref{CNE1}).
Nonetheless, the radial part of the kinetic term $g^{ab}(\nabla_{a}\Psi^*) (\nabla_{b}\Psi)$ is 
$K_r:=g^{rr}R'^2 S^2(\theta)= \Delta R'^2 S^2(\theta)/\rho^2= (r-M)^2 R'^{2} S^2(\theta)/\rho^2$, since 
$\Delta= (r-r_H)(r-r_{-})$ which in the extremal case $r_H=M=r_{-}$ becomes $\Delta=(r-M)^2$. Here we omitted
the labels $n,l,m$ for brevity. So at the extremal horizon
$K_r|_H = C^2 S^2(\theta)/\rho^2$, which is finite, with $\rho^2= M^2(1+\cos^2\theta)$. Thus the kinetic scalar 
remains finite at the horizon despite the unboundedness of the derivative of the radial function at the extremal horizon. Moreover, the radial 
Eq.~(\ref{radialE}) is also satisfied at $r_H=M$ despite that regularity conditions Eq.(\ref{CNE1}) and Eq.(\ref{CNE2}) diverge at 
$r_H=M$. This is by virtue of the factor $\Delta=(r-M)^2$, which makes the terms with the derivatives of $R_{nlm}$ to vanish at $r=M$.
Furthermore the coefficient of the term with $R_{nlm}$ also vanishes at the horizon given that the quantity ${\cal H}\equiv 0$.

\begin{figure}[h]
\begin{center}
\includegraphics[width=0.5\textwidth]{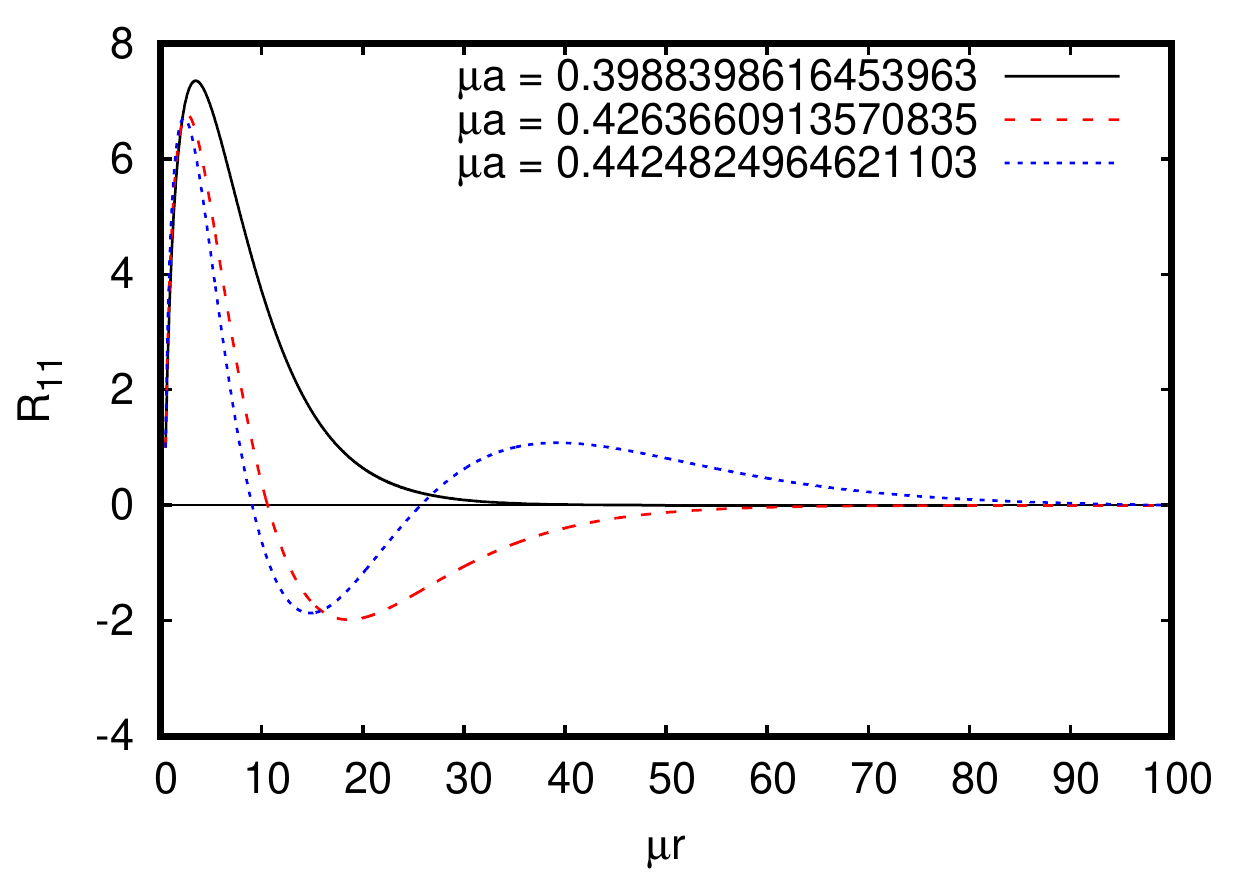}
\includegraphics[width=0.5\textwidth]{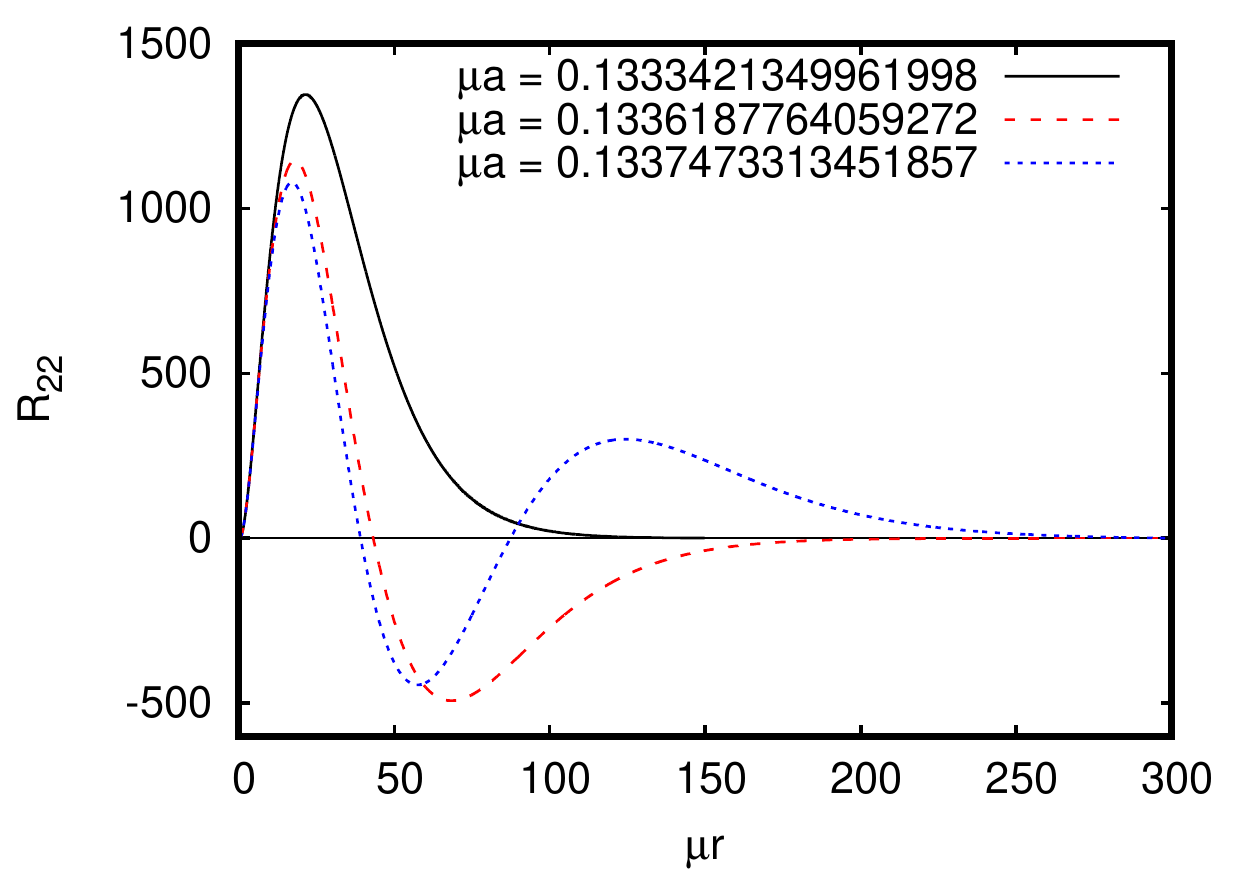}
\caption{(color online) Radial Solutions $R_{n11}$ ($l=1=m$) (top panel) and $R_{n22}$ $l=2=m$ (bottom panel) with principal numbers 
$n = 0,1,2$ (number of nodes) in a Kerr background with the horizon located at $\mu r_H = 0.5$. The corresponding eigenvalues $a_{nlm}$ 
are displayed.}\label{fig:R11r05}
\end{center}
\end{figure}

Figure~\ref{fig:R11r05} depicts
some examples of radial solutions with different nodes with a fixed value $\mu r_H=0.5$: $R_{n11}$ ($l=m=1$) and $R_{n22}$ ($l=m=2$) with $n=0,1,2$.

\begin{figure}[h]
\begin{center}
\includegraphics[width=0.31\textwidth]{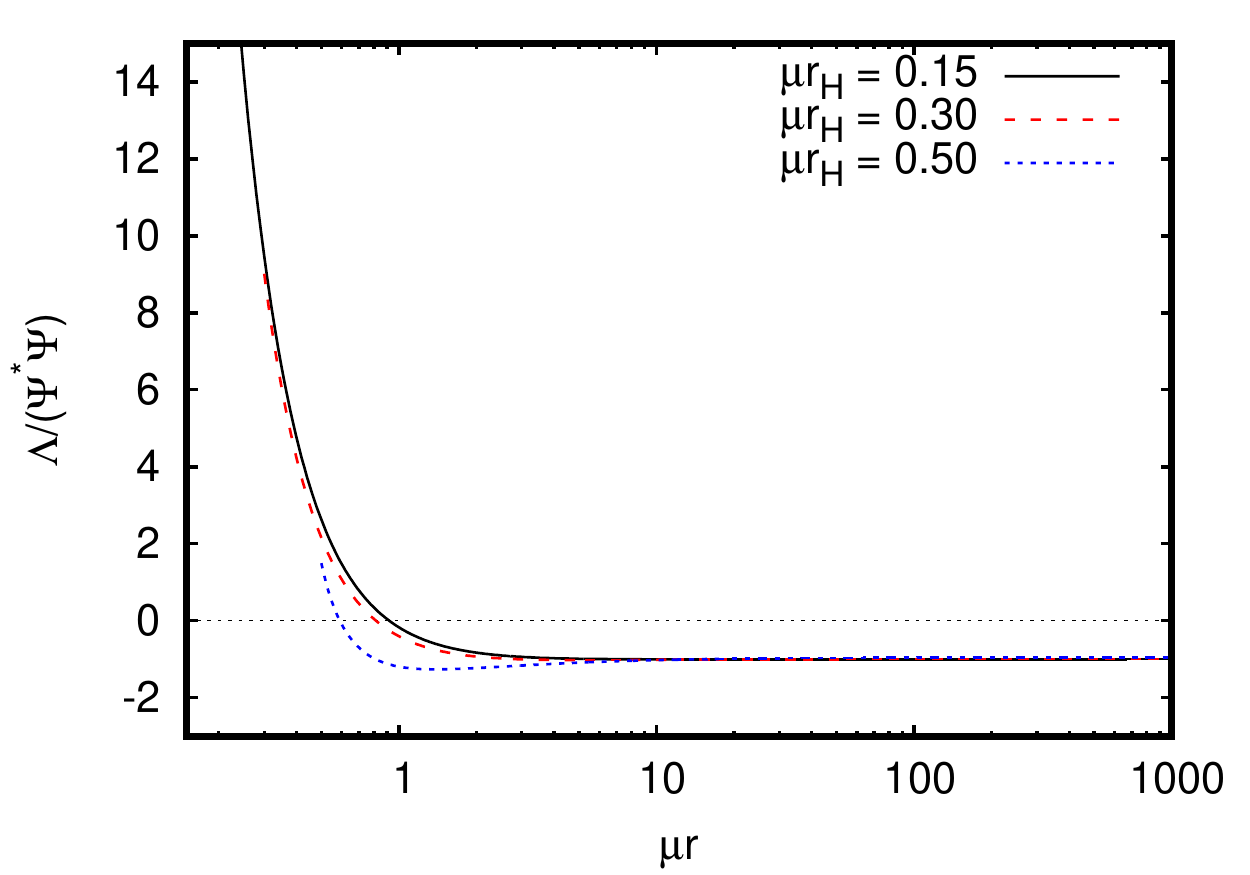}
\includegraphics[width=0.31\textwidth]{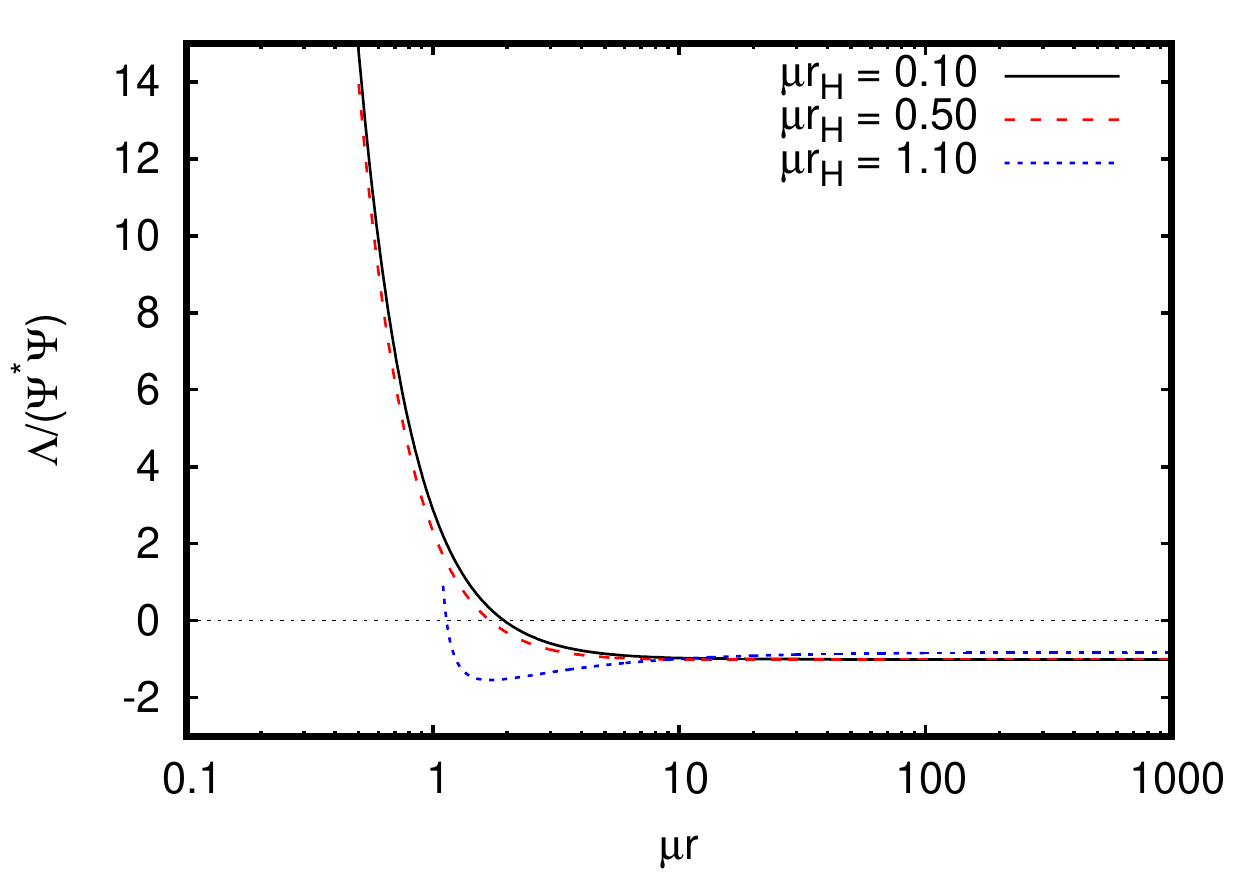}
\includegraphics[width=0.31\textwidth]{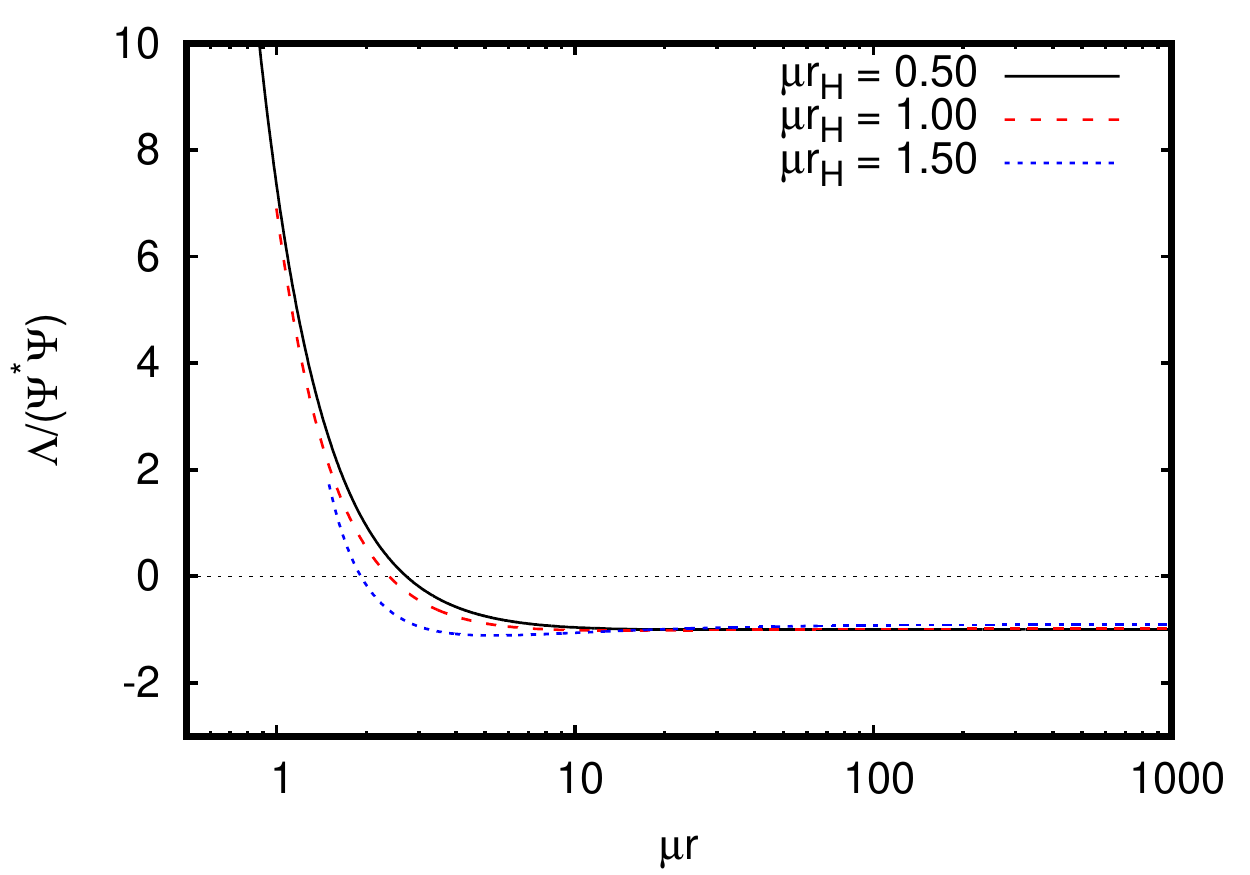}
\caption{(color online) Rotational contribution $\Lambda/\Psi^*\Psi= m^2{\cal R}$ given by Eq.~(\ref{Ibrack}) 
to the total kinetic term $K=(\nabla_c \Psi^*) (\nabla^c\Psi)$ given by Eq.~(\ref{kinstataxi}) 
computed from the {\it regular} cloud solutions in the Kerr background at $\theta=\pi/2$ for different horizon locations $r_H$ and for 
configurations with ``quantum numbers'' $n = 0$ and $m = l = 1$ (top panel), $m = l = 2$ (middle panel), and $m = l = 3$ (bottom panel). 
The panels show that in all the cases $\Lambda/\Psi^*\Psi$ has a negative contribution in most of the BH's domain of outer communication 
which indicates that the four-dimensional volume integral of the total kinetic term $K$ may vanish 
despite the fact that non-trivial cloud configurations exist. Asymptotically $\Lambda/\Psi^*\Psi\rightarrow -m^2\Omega_H^2$ 
[see Eq.~(\ref{Ibrack}) and Table~\ref{tab:kin}].}\label{fig:kinetic}
\end{center}
\end{figure}

Figure~\ref{fig:kinetic} plots the rotational part of the kinetic term Eq.~(\ref{kinstataxi}) that appears in the integral (\ref{intident1})
for some of the numerical solutions presented above and evaluated at $\theta=\pi/2$, for simplicity. At the horizon this quantity is positive but soon becomes negative. Here 
we normalized this rotational part to the square of the cloud amplitude $\Psi^*\Psi$, which is positive, to better appreciate the 
positive and negative values. Due to the use of this normalization, this quantity does not vanish asymptotically. In fact it reaches the 
{\it negative} constant value given by Eq.~(\ref{Lambdaasym}) $\Lambda/\Psi^*\Psi \rightarrow -m^2\Omega_H^2$, where $\Psi^*\Psi=\phi^2$ 
(see Table~\ref{tab:kin} for the numerical values of that quantity at the horizon and asymptotically). This also corroborates that our numerical results are consistent with the analytic expectations. 
The fact that the rotational contribution is negative in most part of the domain of outer communication of the Kerr BH 
indicates that the integral vanishes precisely due to the presence of such negative contribution without the need for the field configuration 
$\Psi(t,r,\theta,\phi)$ to vanish identically, something that is required in the non-rotating situation 
(i.e. the spherically symmetric scenario). These results corroborates our initial expectations concerning the inequality (\ref{haircond}) which
allows us to understand in simple grounds the existence of the non-trivial boson clouds.

Finally, it is worth mentioning the following intriguing scenario that results from the numerical analysis. 
We have checked numerically that in the limit $r_H\rightarrow 0$ one finds cloud configurations with
  $a\rightarrow 0$, $M\rightarrow 0$, $w/\mu\rightarrow 1$ and $\Omega_H/\mu \rightarrow 1/m$
(cf. Tables~\ref{tab:spectrum1}--\ref{tab:spectrum3}). One can understand this behavior from analytic expressions as follows. 
From Eq.~(\ref{fluxcond2}) we see that taking $w/\mu= 1$ leads to $\Omega_H/\mu = 1/m$. Thus, 
Eqs.~(\ref{OmH}) and (\ref{rH}) yield $r_H= 2M/[1+(2M\mu/m)^2]$ and $a= (m/\mu) (2M\mu/m)^2 /[1+ (2M\mu/m)^2]$. 
Alternatively $r_H= \frac{m}{\mu} \sqrt{\frac{a\mu}{m}(1-\frac{a\mu}{m})}$ and $M=\frac{m}{2\mu}\sqrt{\frac{a\mu/m}{1-a\mu/m}}$. So, for $a\mu/m\ll 1$, 
we have $r_H\approx \frac{1}{\mu} \sqrt{ma\mu}$ and $M\approx \frac{1}{2\mu}\sqrt{ma\mu}\approx r_H/2$. 
These expressions reproduce very well the numerical values displayed in the first rows of 
Tables~\ref{tab:spectrum1}--\ref{tab:spectrum3}. Furthermore, since asymptotically $R(r) \sim e^{-r\sqrt{\mu^2-\omega^2}}/r$,
in the limit $w/\mu\rightarrow 1$, $R(r)\sim 1/r$, and the gradients of $R(r)$
become steeper at the horizon ($R'_{r_H\rightarrow 0}\rightarrow \infty$) and the maximum amplitude grows (cf. Figure~\ref{fig:R11}).  
In principle when $r_H=0=a=M$ the BH ``disappears'', and what remains is the Minkowski spacetime
 and the clouds become singular at the origin $r_H=0$ since 
$R'_{r_H\rightarrow 0}\rightarrow \infty$. If one tries to construct cloud configurations in Minkowski spacetime 
that are regular at the origin and well behaved asymptotically then the only possible configurations are with $l=0=m$. 
Among these the configurations that are well behaved asymptotically are:
a) The trivial solution $R(r)= const$ for $\omega=\mu$, but only $R(r)\equiv 0$ has finite energy; 
b) The solution $R(r)= R_0 \sin(\lambda r)/(\lambda r)$, for $\omega^2>\mu^2$, where $R_0$ 
is a constant and $\lambda=\sqrt{\omega^2-\mu^2}$. This solution is not localized either and has infinite energy. 
Thus, the only solution that is well behaved everywhere and has finite energy is the trivial one. This is in agreement 
with the integral analysis similar to the one performed in Section~\ref{sec:obstructions}: if one considers Minkowski 
spacetime and possible cloud solutions in spherical symmetry, then the integral in the inner boundary (the origin at $r=0$) 
vanishes (assuming that the normal at the origin has component only in the radial direction) 
by regularity $R'|_{r=0}=0$, and the surface integral at infinity 
vanishes as well if one demands that the field is localized, i.e., has finite energy. Then, if $\omega\neq 0$, one has 
$\int_{\cal V} \Big[ \phi^2 \Big(\mu^2 - \omega^2\Big) +  (\partial_r \phi)^2 \Big] \sqrt{-g}d^4 x \equiv 0$. 
Localized solutions require $\mu^2>\omega^2$. Thus the terms involving $\phi^2$ and $(\partial_r \phi)^2$ are 
non-negative, and each has to vanish separately. Thus, we conclude that the only possible regular solution is the trivial one 
$\phi(r)\equiv 0$.

In the full Einstein-boson-field system the boson field is not a test-field anymore, and so when 
$r_H=0$, the configuration becomes a globally regular boson star, rotating or not. In this case there is
a balance between gravity, rotation and the effective pressure gradients associated with the boson field. In particular, 
in the absence of rotation, boson stars can also exist which are not prevented
by {\it no-go theorems}. In other words, only for finite $r_H$, i.e. $r_H\neq 0$ is that the no-hair theorems
apply in the non-rotating case. A similar argument to the above one, but for globally regular self-gravitating spherically symmetric boson stars, 
leads to $\int_{\cal V} \Big[ \frac{\phi^2}{N^2}\Big(N^2\mu^2 - \omega^2\Big) +  g^{rr}(\partial_r \phi)^2 \Big] \sqrt{-g}d^4 x \equiv 0$, 
where we used $g^{tt}=-1/N^2$. For an asymptotically flat spacetime $0<N<1$. Thus, even if $\mu^2>\omega^2$, which is required for 
localized solutions, the quantity $N^2\mu^2 - \omega^2$ can be negative, notably near the origin. Thus, the above integral can vanish 
without requiring $\phi(r)\equiv 0$.

\section{Conclusion}
The integral technique used to prove the no-hair theorem for a complex-valued (boson) scalar field in the spherically symmetric 
scenario has been employed to understand in a heuristic fashion the existence of non-trivial hair (boson clouds) in the stationary and axisymmetric 
spacetime of a Kerr BH. In particular, we show that when non-trivial boson clouds exist the integral can vanish due to the presence of negative terms 
that compensate the positive ones. Such negative terms are absent in the non-rotating situation and thus the only way that such 
an integral can vanish is by the absence of hair (i.e. when the boson field is identically null in the domain of outer communication 
of the BH). In view of this we conclude that spacetimes containing a BH with less symmetries than the spherical and static 
scenarios pose serious obstacles towards a generalization of the no-hair theorems. Moreover, the fact that regular cloud configurations maybe 
supported by a Kerr BH, and more generically, by rotating BH's that are not necessarily Kerr (black holes with boson hair)~\cite{Herdeiro2014,Herdeiro2015}
shows that rotating boson stars if collapsed might produce a new 
kind of black hole endowed with new types of numbers (hair) other than the two parameters $M$ and $J$ ($J=aM$).
Alternatively, it is also possible that such boson hair may form due to the development of superradiance instabilities
in a Kerr BH~\cite{superradiance} (see \cite{Brito2015} for a review).

\bigskip
\section*{Acknowledgments}
This work was supported partially by DGAPA--UNAM grants IN107113, IN111719 and SEP--CONACYT grants CB--166656. G.G. acknowledges CONACYT scholarship 
291036. We are indebted to J. C. Degollado, C. Herdeiro and S. Hod for fruitful discussions and valuable suggestions.


\clearpage


\begin{table}[htbp]
\begin{center}
\caption{Precise eigenvalues $\mu a$ found for different values of $\mu r_{H}$ that allow for stationary configurations 
of a massive scalar field (boson clouds) around a Kerr black hole. The numerical data corresponds to the {\it fundamental mode} 
with numbers $n = 0$ (zero nodes) and $l = m = 1$. Notice the eigenvalue for the near extremal case $a\approx r_H \approx M$ (the last row in the table).
Figure~\ref{fig:spectrum1} shows these eigenvalues pictorially.}
\begin{tabular}{|c|c|c|c|}
\hline 
$\mu r_{H}$ & $\mu a$ & $\Omega_{H}/\mu$ & $\mu M$\\
\hline \hline 
0.05 &  0.0025060835912 &  0.99992145182803427  & 0.02506280454966172 \\ \hline
0.06 &  0.0036126413700 &  0.99988657574844975  & 0.03010875981390454 \\ \hline                     
0.07 &  0.0049234783425 &  0.99984519513851144  & 0.03517314742135750 \\ \hline
0.08 &  0.0064401695538 &  0.99979721572713920  & 0.04025922364926862 \\ \hline
0.09 &  0.0081645543739 &  0.99974217437868562  & 0.04537033304513992 \\ \hline
0.10 &  0.0100987546410 &  0.99968024063106697  & 0.05050992422649560 \\ \hline
0.12 &  0.0146065524928 &  0.99953474856271751  & 0.06088896406554481 \\ \hline
0.15 &  0.0230125053546 &  0.99925877712476252  & 0.07676525134231830 \\ \hline
0.17 &  0.0297566518588 &  0.99903286796104240  & 0.08760428920542891 \\ \hline
0.20 &  0.0416797488169 &  0.99862346683626935  & 0.10434300365418323 \\ \hline
0.22 &  0.0509044390801 &  0.99829729968225567  & 0.11588923163197672 \\ \hline
0.24 &  0.0612205143171 &  0.99792271700668223  & 0.12780823202763017 \\ \hline
0.26 &  0.0727030002659 &  0.99749297988666275  & 0.14016485816861152 \\ \hline
0.28 &  0.0854434480278 &  0.99699961453513208  & 0.15303675501945793 \\ \hline
0.30 &  0.0995546154238 &  0.99643176832449687  & 0.16651853575366765 \\ \hline
0.32 &  0.1151771122918 &  0.99577528497975087  & 0.18072776124359513 \\ \hline
0.34 &  0.1324891052699 &  0.99501133332119163  & 0.19581376914006629 \\ \hline
0.36 &  0.1517209649688 &  0.99411420096045600  & 0.21197118223763078 \\ \hline
0.38 &  0.1731783024484 &  0.99304766928203991  & 0.22946147952492649 \\ \hline
0.40 &  0.1972800571230 &  0.99175865379227768  & 0.24864927617311691 \\ \hline
0.42 &  0.2246255692082 &  0.99016525559278612  & 0.27006743612161038 \\ \hline
0.44 &  0.2561227150420 &  0.98813216117327607  & 0.29454414222788100 \\ \hline
0.46 &  0.2932609481730 &  0.98541328422621710  & 0.32348041709063452 \\ \hline
0.48 &  0.3387899286324 &  0.98149164861627591  & 0.35956105806542960 \\ \hline
0.50 &  0.3988398616453 &  0.97498400601548563  & 0.40907323523731887 \\ \hline
0.51 &  0.4389229798645 &  0.96945268013285657  & 0.44387586495413628 \\ \hline
0.52 &  0.4913899147443 &  0.96000083687141713  & 0.49217696953125500 \\ \hline
0.521 &  0.4975002203326 &   0.95867163794022692 & 0.49803020079758353 \\ \hline
0.522 &  0.5037474862440 &  0.95724800995925607  & 0.50406659951834309 \\ \hline
0.523 &  0.5101038696898 &  0.95572505073135816  & 0.51026286603500048 \\ \hline
0.524 &  0.5165260631493 &  0.95410001795311583  & 0.51657936442043684 \\ \hline
0.525 &  0.5229513465680 &  0.95237367301106302  & 0.52295534369275087 \\ \hline
\end{tabular}
\label{tab:spectrum1}
\end{center}
\end{table}

\begin{table}[htbp]
\begin{center}
\begin{tabular}{|c|c|c|c|}
\hline 
$\mu r_{H}$ & $\mu a$ & $\Omega_{H}/\mu$ & $\mu M$\\
\hline \hline 
0.05 &  0.0012507388607 &  0.49998268561237552 &  0.02501564347697901 \\ \hline
0.06 &  0.0018015325288 &  0.49997495843400130 &  0.03002704599543651 \\ \hline
0.07 &  0.0024528408250 &  0.49996588087563448 &  0.03504297448652243 \\ \hline
0.08 &  0.0032048495464 &  0.49995538662081246 &  0.04006419412884629 \\ \hline
0.09 &  0.0040577739998 &  0.49994347723599086 &  0.04509147516574370 \\ \hline
0.10 &  0.0050118590035 &  0.49993013930549052 &  0.05012559365335877 \\ \hline
0.12 &  0.0072246400129 &  0.49989913652451917 &  0.06021748093048933 \\ \hline
0.15 &  0.0113103786077 &  0.49984162381593678 &  0.07542641554750320 \\ \hline
0.18 &  0.0163257704413 &  0.49977056183280183 &  0.09074036327917228 \\ \hline
0.20 &  0.0201923667214 &  0.49971542288989229 &  0.10101932918452877 \\ \hline
0.23 &  0.0267884708130 &  0.49962064204963164 &  0.11656004819238902 \\ \hline
0.26 &  0.0343565360814 &  0.49951076588295362 &  0.13226994532983188 \\ \hline
0.29 &  0.0429181276274 &  0.49938498949136612 &  0.14817580289491444 \\ \hline
0.32 &  0.0524983687660 &  0.49924235863092015 &  0.16430637300484596 \\ \hline
0.35 &  0.0631263215073 &  0.49908175049895254 &  0.18069276066722231 \\ \hline
0.38 &  0.0748354483949 &  0.49890184403263105 &  0.19736887412694826 \\ \hline
0.41 &  0.0876641747179 &  0.49870108919012102 &  0.21437196040120488 \\ \hline
0.44 &  0.1016565730867 &  0.49847766311902936 &  0.23174324869517260 \\ \hline
0.47 &  0.1168632016269 &  0.49822941840883611 &  0.24952873180267321 \\ \hline
0.50 &  0.1333421349961 &  0.49795381570404657 &  0.26778012496534481 \\ \hline
0.53 &  0.1511602415373 &  0.49764783821919800 &  0.28655605530343875 \\ \hline
0.56 &  0.1703947773934 &  0.49730788052398117 &  0.30592355371694674 \\ \hline
0.59 &  0.1911353943915 &  0.49692960327267016 &  0.32595994829595965 \\ \hline
0.62 &  0.2134866945864 &  0.49650774040886420 &  0.34675529739150768 \\ \hline
0.65 &  0.2375715169819 &  0.49603583963768239 &  0.36841555821625277 \\ \hline
0.68 &  0.2635352191449 &  0.49550590754551932 &  0.39106677333073625 \\ \hline
0.71 &  0.2915513302711 &  0.49490791422719244 &  0.41486068886120059 \\ \hline
0.74 &  0.3218291287799 &  0.49422909183391317 &  0.43998242441303426 \\ \hline
0.77 &  0.3546239790917 &  0.49345293158740516 &  0.46666114710837614 \\ \hline
0.80 &  0.3902514790233 &  0.49255754439949428 &  0.49518513554996796 \\ \hline
0.83 &  0.4291077554304 &  0.49151352411414989 &  0.52592377456060224 \\ \hline 
0.86 &  0.4716978737846 &  0.49028003416693378 &  0.55935981635639864 \\ \hline
0.89 &  0.5186770447682 &  0.48879879015609956 &  0.59613813301661600 \\ \hline
0.92 &  0.5709083191018 &  0.48698339785846745 &  0.63713929827160543 \\ \hline
0.95 &  0.6295398490509 &  0.48470122474901445 &  0.68358969554902760 \\ \hline
0.98 &  0.6960768909483 &  0.48173976923901518 &  0.73720563168999498 \\ \hline
1.01 &  0.7723176929090 &  0.47774948581819660 &  0.80028446474278925 \\ \hline
1.04 &  0.8595831399802 &  0.47217307580898582 &  0.87523229545114767 \\ \hline
1.07 &  0.9556393260788 &  0.46432035623969281 &  0.96175071100398757 \\ \hline
1.10 &  1.0494928241257 &  0.45404380344452971 &  1.05065235813250960 \\ \hline
1.13 &  1.1251833481143 &  0.44247383922115080 &  1.12519361366102790 \\ \hline
1.131 &  1.1272390804412 & 0.44208419662197024 &   1.1272453335429389 \\ \hline 
1.132 &  1.1292545637580  & 0.44169481087704149 &  1.1292578930072801 \\ \hline
1.133 &  1.1312273567029  & 0.44130572557988013 &  1.1312287434038142 \\ \hline
1.134 &  1.1331507640318  & 0.44091698385154393 &  1.1331510820220714 \\ \hline
\end{tabular}
\caption{Similar to Table~\ref{tab:spectrum1} but for the mode $l = m = 2$ with $n = 0$. These eigenvalues are associated with the dotted line marked 
$m=2$ in Figure~\ref{fig:MvsOm}. Figure~\ref{fig:spectrum2} shows these eigenvalues pictorially.}
\label{tabla:spectrum2}
\end{center}
\end{table}

\begin{table}[htbp]
\begin{center}
\begin{tabular}{|c|c|c|c|}
\hline 
$\mu r_{H}$ & $\mu a$ & $\Omega_{H}/\mu$ & $\mu M$\\
\hline \hline 
0.05 &  0.0008335486538  & 0.33332682290664223 &  0.02500694803358356 \\ \hline
0.06 &  0.0012004465836  & 0.33332395550806182 &  0.03001200893333494 \\ \hline
0.07 &  0.0016341608543  & 0.33332055671945948 &  0.03501907486927115 \\ \hline
0.08 &  0.0021347454793  & 0.33331664207791856 &  0.04002848211413427 \\ \hline
0.09 &  0.0027022627997  & 0.33331220731244143 &  0.04504056791243719 \\ \hline
0.10 &  0.0033367834670  & 0.33330723829324627 &  0.05005567061952897 \\ \hline
0.15 &  0.0075175078711  & 0.33327438049768110 &  0.07518837641530841 \\ \hline
0.20 &  0.01338885596674027  & 0.33322802557492959 &  0.10044815366024530 \\ \hline
0.25 &  0.02096948579087086  & 0.33316776770931733 &  0.12587943866866708 \\ \hline
0.30 &  0.03028386215652325  & 0.33309309312668517 &  0.15152852051185883 \\ \hline 
0.35 &  0.04136263503158459  & 0.33300335017807181 &  0.17744409653822293 \\ \hline
0.40 &  0.05424312789118017  & 0.33289773286256774 &  0.20367789615427367 \\ \hline
0.45 &  0.06896995391843294  & 0.33277526125898832 &  0.23028539393723418 \\ \hline
0.50 &  0.08559578331167261  & 0.33263475533190112 &  0.25732663812073880 \\ \hline
0.55 &  0.10418229243354787  & 0.33247480178467015 &  0.28486722732428121 \\ \hline
0.60 &  0.12480133500042097  & 0.33229371226110288 &  0.31297947768157275 \\ \hline
0.65 &  0.14753638775345154  & 0.33208946984260845 &  0.34174383516256679 \\ \hline
0.70 &  0.1724843405273466  & 0.33185966176219373 &  0.37125060551939548 \\ \hline
0.75 &  0.1997577174973116 &  0.33160138509117232 &  0.40160209713315714 \\ \hline
0.80 &  0.2294874657785451 &  0.33131114239176879 &  0.43291531059341193 \\ \hline   
0.85 &  0.2618264539908034 &  0.33098467932882158 &  0.46532534824082250 \\ \hline
0.90 &  0.2969539275563022 &  0.33061678835837982 &  0.49898979727284087 \\ \hline
0.95 &  0.3350811891408546 &  0.33020101516240069 &  0.53409442279792063 \\ \hline
1.00 &  0.3764590344232502 &  0.32972933055898107 &  0.57086070229944297 \\ \hline
1.05 &  0.4213872455033129 &  0.32919150025084232 &  0.60955581460612829 \\ \hline
1.10 &  0.4702274756275244 &  0.32857446397648132 &  0.65050630856137925 \\ \hline
1.15 &  0.5234199895332505 &  0.32786114746763201 &  0.69411673280129915 \\ \hline
1.20 &  0.5815059153458475 &  0.32702876585074520 &  0.74089547065925498 \\ \hline
1.25 &  0.6451573899937433 &  0.32604651631125553 &  0.79149122314541565 \\ \hline
1.30 &  0.7152157742625678 &  0.32487161360744093 &  0.84674369375154013 \\ \hline
1.35 &  0.7927354534159784 &  0.32344278107804569 &  0.90775166633430993 \\ \hline
1.40 &  0.87902763896390812 & 0.32167124419634047 &  0.97596056065810255 \\ \hline
1.45 &  0.97564752155589107 & 0.31942485689859007 &  1.0532372841918607 \\ \hline
1.50 &  1.0841840817873019  & 0.31650803843378561 &  1.1418183744003250 \\ \hline
1.55 &  1.2054512147772587  & 0.31264842453833058 &  1.2437460100670867 \\ \hline
1.60 &  1.3373971374311400  & 0.30754439861413574 &  1.3589472197528147 \\ \hline
1.65 &  1.4722454635549960  & 0.30107227911614109 &  1.4818202136237166 \\ \hline
1.70 &  1.5977145896069918  & 0.29355227451282800 &  1.6007917381891290 \\ \hline
1.75 &  1.7046041031612489  & 0.28561562969304383 &  1.7051928995754762 \\ \hline
1.80 &  1.7903199625769979  & 0.27777373935202326 &  1.7903459912226953 \\ \hline
1.81 &  1.8048562172163640  & 0.27624197524817018 &  1.8048635261946311 \\ \hline
1.82 &  1.8183980131169923  & 0.27472516820653298 &  1.8183987181614907 \\ \hline
1.821 &  1.819680931489054  & 0.27457433757760219 &  1.8196814092325291 \\ \hline
1.822 &  1.820952923898013  & 0.27442366486653114 &  1.8209532247674876 \\ \hline
1.823 &  1.822218850860257  & 0.27427315089303994 &  1.8222190182200986 \\ \hline
1.824 &  1.823472742675874  & 0.27412279556145724 &  1.8234728188820932 \\ \hline
1.825 &  1.824695526557144  & 0.27397259892623527 &  1.8246955519555224 \\ \hline

\end{tabular}
\caption{Similar to Table~\ref{tab:spectrum1} but for the mode $l = m = 3$ with $n = 0$. These eigenvalues are associated with the dotted line marked 
$m=3$ in Figure~\ref{fig:MvsOm}. Figure~\ref{fig:spectrum3} shows these eigenvalues pictorially.}
\label{tab:spectrum3}
\end{center}
\end{table}

\begin{table}[h]
\centering
\begin{tabular}{|c|c|c|c|}
\hline
\multicolumn{4}{|c|}{Kinetic term} \\
\hline
 $m$ & $r_H$ & $\dfrac{\Lambda}{\Psi^*\Psi }\left(r = r_{H}\right)$ & $\dfrac{\Lambda}{\Psi^*\Psi }\left(r \rightarrow \infty \right)$\\
\hline \hline
\multirow{3}{0.2cm}{1} & 0.15 & 42.423903339347923 & -0.99867038923580298\\ \cline{2-4}
& 0.30 & 9.0160191201301796 & -0.99320604286679159\\ \cline{2-4}
& 0.50 &  1.4939562556780108 & -0.95137114225247021 \\ \hline
\multirow{3}{0.2cm}{2} & 0.10 & 397.99802339758315 & -0.99981682617106094\\ \cline{2-4}
& 0.50 & 13.945796372744445 &  -0.99235942720910209\\ \cline{2-4}
& 1.10 & 0.90590341906419636 & -0.82635553483390189\\ \hline
\multirow{3}{0.2cm}{3} & 0.50 & 33.979191729610299 &  -0.99631663635406698\\ \cline{2-4}
& 1.00 & 6.9043412931258255 & -0.97960225256054256\\ \cline{2-4}
& 1.50 & 1.7257919562766677 & -0.90365058480094529\\ \hline
\end{tabular}
\caption{Values of rotational contribution to the kinetic term at $r = r_{H}$ and $r \rightarrow \infty$.}
\label{tab:kin}
\end{table}

\end{document}